\documentclass[10pt,journal]{IEEEtran}

\usepackage{multirow}
\usepackage{booktabs}
\usepackage{xcolor,colortbl, multirow }
\usepackage{tabularray}
\usepackage{graphicx}
\usepackage{amssymb}
\usepackage[nointegrals]{wasysym}
\usepackage{amsmath}
\usepackage{caption}
\def\DOWNcircle{\rotatebox[origin=c]{90}{\LEFTcircle}}
\usepackage{array}
\usepackage{makecell}

\newcommand{\minew}[1]{{\color{black}{#1}}}

\ifCLASSOPTIONcompsoc
  \usepackage[nocompress]{cite}
\else
  \usepackage{cite}
\fi

\ifCLASSINFOpdf
\else
\fi

\begin{document}

\title{A Survey on Semantic Communication for Vision: Categories, Frameworks, Enabling Techniques, and Applications}


\author{Runze~Cheng,~\IEEEmembership{Member,~IEEE,}
        Yao~Sun,~\IEEEmembership{Senior Member,~IEEE,}
        Ahmad~Taha,~\IEEEmembership{Senior Member,~IEEE,}
        Xuesong~Liu, ~\IEEEmembership{Student Member,~IEEE,}        David Flynn, ~\IEEEmembership{Senior Member,~IEEE,}
and~Muhammad~Ali~Imran,~\IEEEmembership{Fellow,~IEEE}
\thanks{This work was supported in parts by the Engineering and Physical Sciences Research Council (EPSRC) grant ``Transit" (EP/Y006909/1)}
\thanks{Runze Cheng, Yao Sun, Ahmad Taha, Xuesong Liu, David Flynn, and Muhammad Ali Imran are with the James Watt School of Engineering, University of Glasgow, Glasgow, G12 8QQ, UK. 
}
\thanks{Yao Sun is the corresponding author. Email: Yao.Sun@glasgow.ac.uk}

}

\markboth{Journal of \LaTeX\ Class Files,~Vol.~14, No.~8, August~2015}%
{Shell \MakeLowercase{\textit{et al.}}: Bare Demo of IEEEtran.cls for Computer Society Journals}

\IEEEtitleabstractindextext{%
\begin{abstract}
Semantic communication (SemCom) emerges as a transformative paradigm for traffic-intensive visual data transmission, shifting focus from raw data to meaningful content transmission and relieving the increasing pressure on communication resources. \minew{However, to achieve SemCom, challenges are faced in accurate semantic quantization for visual data, robust semantic extraction and reconstruction under diverse tasks and goals, transceiver coordination with effective knowledge utilization, and adaptation to unpredictable wireless communication environments. }
In this paper, we present a systematic review of SemCom for visual data transmission (SemCom-Vision), wherein an interdisciplinary analysis integrating computer vision (CV) and communication engineering is conducted to provide comprehensive guidelines for the machine learning (ML)-empowered SemCom-Vision design. Specifically, this survey first elucidates the basics and key concepts of SemCom. Then, we introduce a novel classification perspective to categorize existing SemCom-Vision approaches as semantic preservation communication (SPC), semantic expansion communication (SEC), and semantic refinement communication (SRC) based on communication goals interpreted through semantic quantization schemes.\minew{ Moreover, this survey articulates the ML-based encoder-decoder models and training algorithms for each SemCom-Vision category, followed by knowledge structure and utilization strategies.} Finally, we discuss potential SemCom-Vision applications.

\end{abstract}

\begin{IEEEkeywords}
Semantic Communication, Machine Learning, Visual Data Transmission, Transceiver Framework, Knowledge Structure and Utilization
\end{IEEEkeywords}}

\maketitle

\IEEEdisplaynontitleabstractindextext

\IEEEpeerreviewmaketitle

\section{Introduction}\label{sec:introduction}

\IEEEPARstart{V}{isual} data has become the most dominant source of communication load, accounting for more than 82.5\% of global network traffic \cite{erdem2025synthesis}. As visual content continues to rapidly expand in both volume and quality, significant burdens can be posed on constrained communication resources, especially spectrum \cite{park2024transmit, zaffar2021vpr}. In this context, alleviating the resource burdens while ensuring effective visual data delivery has become a critical research focus.


\subsection{Background}
Semantic communication (SemCom), as a revolutionary paradigm, offers a promising path for visual data transmission by fundamentally shifting the focus from transmitting raw data to conveying the intended meaning of information \cite{iyer2023survey, yang2022semantic, lu2023semantics, chaccour2024less}. In SemCom, only the meaningful semantics of visual content are selectively extracted and transmitted to alleviate communication resource burdens, and then reconstructed as visual content in a semantic-aware manner to preserve the essential quality \cite{zhang2024intellicise, getu2024survey, getu2025semantic}.

To enable SemCom potential for visual data transmission (SemCom-Vision), it is essential to first understand the nature of visual data. Typically, visual data is captured by one or more sensors, including traditional cameras, remote sensing instruments, X-ray imaging systems, radar, or ultrasound receivers. Depending on the sensors, the resulting content can range from standard 2D photographs to 3D reconstructions, time-sequenced videos, or multispectral datasets. The pixel values in these visual data may represent optical intensity, depth, reflectivity, or even non-visible phenomena such as electromagnetic absorption or acoustic echoes \cite{zhao2024review}. Since these visual data collectively represent content that allows humans and intelligent entities to perceive, interpret, and understand the world visually, they can be comprehensively referred to as \textit{images} in the broadest sense \cite{wang2024visual}.
From a conventional communication level, the information of images is all about the pixel value, pixel position (variation and distribution), which is relatively intuitive and with well-defined regulations in transmission \cite{yuan2024survey}. However, in SemCom, the focus shifts to the image objects, their corresponding structural relationships, spatial arrangements, or even to higher semantically aspects such as emotion, atmosphere, intent, and contextual meaning \cite{hosonuma2024image}.
In this case, to effectively transmit the semantics of images, SemCom-Vision faces significant challenges:

\textit{Challenge 1 semantic quantization:} To exploit SemCom-Vision, it is vital to identify and quantify the semantics in images \cite{qiu2024semantic}. Nevertheless, even images with modest resolution can contain rich spatial, temporal, and contextual information distributed across different scales and regions. Moreover, from diverse tasks, the fundamentals of what constitutes the ``meaning" of images are different \cite{wang2024feature}. For instance, a medical X-ray image may contain critical diagnostic information in subtle intensity variations that are imperceptible to untrained observers, while a surveillance image may prioritize motion patterns and object identification. Furthermore, even the same image can carry different semantic significance depending on the tasks and goals of SemComs \cite{ma2023task}. Quantifying these multi-faceted semantics in a unified, measurable framework that can guide semantic processing remains an open and challenging research problem \cite{fu2023vector}.

\textit{Challenge 2 semantic extraction and reconstruction:} Semantic extraction for visual data involves sophisticated perception and reasoning capabilities that adapt to diverse tasks and goals \cite{wei2023federated}. Unlike conventional compression techniques that apply uniform processing across entire images, semantic extraction requires developing a robust algorithm that can extract meaningful semantic features while discarding irrelevant information \cite{cheng2025semantic}. Furthermore, reconstructing meaningful semantic features might require accurately recovering image details from semantics according to diverse tasks and goals, or even from insufficient or polluted semantics, which is technically complicated \cite{zhao2022background}. 

\minew{\textit{Challenge 3 transceiver coordination with knowledge utilization:} Achieving coordinated semantic understanding among transceivers is inherently difficult, as transmitted visual semantics are often implicit and highly dependent on diverse tasks and goals \cite{liang2023vista, liang2025vista}. Without structured knowledge representation, reliable reasoning, and efficient inference to enable accurate association among objects, properties, and relationships, transceivers may struggle to meet the diverse service requirements of different SemCom-Vision applications.}

\textit{Challenge 4 robustness to communication conditions and requirements:} Wireless environments are often unpredictable, considering varying channel conditions and dynamic resource availability, determining the way of semantic features extraction, transmission, and reconstruction poses additional challenges \cite{xie2021deep}. Moreover, balancing competing service requirements such as latency and reconstruction quality under varying channel conditions and dynamic resource availability also requires complex trade-offs \cite{cheng2024wireless}.

In this context, traditional coding strategies and communication protocols in communication engineering face limitations when addressing the complex and computation-intensive semantic processing, transceivers coordination, and trade-offs between services and resources \cite{nishio2021wireless}. To overcome these challenges, the development of effective SemCom-Vision systems calls for interdisciplinary collaboration, where integrating advances in computer vision (CV) with communication engineering shows potential to meet diverse service requirements while ensuring robustness under varying channel conditions and dynamic resource availability \cite{tian2020applying}. 

   

\subsection{Motivation}
The convergence of machine learning (ML) breakthroughs in CV and the urgent need for efficient visual data transmission create a compelling opportunity for SemCom-Vision development \cite{davies2017computer}. Recent advances in ML have demonstrated unprecedented capabilities in understanding and interpreting visual content, with models achieving human-level or even superhuman performance in tasks such as object recognition, scene understanding, and content generation \cite{mahadevkar2022review}. These ML models have fundamentally transformed the way of representing and processing semantics from visual data, providing the foundational techniques for the SemCom-Vision design \cite{o2020deep}. 

Unlike deterministic rules or statistical models widely used in traditional communication, ML offers a natural synergy with SemCom-Vision, as it enables seamless integration across multiple stages of the communication pipeline. ML models can learn complex mappings between raw visual data and their semantics through large-scale training, enabling more robust and generalizable semantic quantization \cite{sun2024s}. Especially with powerful neural networks (NNs), ML has proven exceptionally effective at capturing hierarchical semantic features across multiple scales and abstraction levels \cite{zhang2024machine}.
In addition, the ML-based encoder-decoder architecture provides a natural backbone for SemCom-Vision. By constructing ML-based semantic encoders that compress visual information into compact semantics and corresponding decoders that reconstruct meaningful visual content from these semantics, it can achieve the fundamental goal of transmitting meaning rather than raw data \cite{rong2025semantic}. These learned semantics can capture essential semantic features while discarding perceptually irrelevant binary bits, leading to significant bandwidth savings without compromising reconstruction quality for intended tasks \cite{niu2025mathematical}.
Furthermore, by analyzing and understanding ML models, adaptive knowledge structures can be developed for different tasks and goals to enhance semantic understanding, which in turn supports ML-based semantic extraction, transmission, and reconstruction \cite{nguyen2021advanced}. Moreover, with ML techniques like continual learning, pre-trained knowledge can be adapted to new user-specific preferences and task-specific requirements \cite{de2021continual}. This adaptability is crucial for addressing the diverse and dynamic service requirements.
The integration of ML with SemCom-Vision also opens new possibilities for intelligent resource allocation and adaptive transmission strategies \cite{won2024resource}. ML models can learn to predict channel conditions, task context, and make service requirements trade-offs, enabling dynamic adjustment of semantic extraction, transmission, and reconstruction parameters \cite{zhang2023toward}. This capability is essential for maintaining robust performance under varying channel conditions and dynamic resource availability.

Given these technical foundations and the outlined pressing challenges, the development of ML-empowered SemCom-Vision systems represents not merely an incremental improvement over existing approaches, but a paradigmatic shift toward more intelligent and efficient visual communication. Specifically, SemCom-Vision encompasses investigation in multiple critical areas, including semantic quantization, transceiver architecture, and knowledge structure. While researchers have proposed various solutions for these components, no widely accepted ML-angle guideline has been established. This gap motivates our comprehensive review of current advances, emerging trends, and future research directions in SemCom-Vision. 

\subsection{Related Surveys}
Several notable surveys have been conducted to comprehensively review SemCom and include the visual data transmission perspective, as summarized in Table \ref{table_1_survey}. The work of \cite{iyer2023survey} offers an in-depth survey on the evolution of SemCom techniques from the standpoint of wireless networks. It outlines potential architectures, identifies current challenges, and highlights future research directions in the SemCom network and SemCom-Vision transceiver design. In \cite{yang2022semantic}, the authors present a holistic review of SemCom advancements and include their practical visual data transmission applications, while categorizing the field into semantic-oriented, goal-oriented, and semantic-aware communication paradigms. The survey of \cite{lu2023semantics} provides a thorough overview of SemCom’s foundational structure and working pipeline, covering its background, enabling technologies, and research taxonomy. It also traces the development of SemCom from theoretical foundations to some practical visual data transmission applications and discusses key challenges for future deployment. In \cite{chaccour2024less}, the authors delve into the core advantages of SemCom and its synergy with existing techniques, emphasizing the paradigm shift from data-driven to knowledge-driven communication. Additionally, the authors in \cite{zhang2024intellicise} present an insightful overview that bridges the conceptual and technical gap between traditional communication models and SemCom, particularly from the aspect of intellicise wireless networks, and include visual data transmission use cases. Furthermore, the works in \cite{getu2024survey} and \cite{getu2025semantic} deliver comprehensive surveys on the SemCom research landscape, including techniques, theoretical underpinnings, and key challenges. These references also provide detailed mathematical formulations of concepts such as semantic information and semantic entropy. Meanwhile, our work covered all aspects from the transceiver design and semantic quantization to encoder-decoder construction, ML-based model training, and knowledge structure and utilization.

\begin{table*}[!t]
\caption{Summary of Related Surveys Versus Our Work}
\label{table_1_survey}
\centering
  \begin{tblr}{
    colspec = {
      Q[c,m,wd=4.82em]      
      Q[l,m,wd=23em]    
      Q[c,m,wd=5em]     
      Q[c,m,wd=5em]
      Q[c,m,wd=5em]
      Q[c,m,wd=5em]
      Q[c,m,wd=5em]
      Q[c,m,wd=5em]
    },
    rowhead = 1,            
    row{1} = {halign=c},    
    stretch = 1.2,          
    row{odd} = {bg=gray9},
  }
\hline
\textbf{References} & \textbf{Contributions} & \textbf{Transceiver Framework}& \textbf{Semantic Quantization} &  \textbf{Encoder-Decoder Construction}  & \textbf{ML-based Model Training} & \textbf{Knowledge Structure and Utilization} \\
\hline
\cite{iyer2023survey} & \parbox[c]{23em}{\textbullet Provides an early overview of SemCom. \\ \textbullet Classifies literature by architecture and applications. \\ \textbullet Discusses high-level challenges and semantic layers.\\ \textbullet Focuses on conceptual taxonomy rather than technical modeling.}   & \DOWNcircle & \DOWNcircle & \Circle & \Circle & \DOWNcircle\\ 

\cite{yang2022semantic}& \parbox[c]{23em}{\textbullet Introduces semantic-, goal-, and semantic-aware frameworks.\\  \textbullet Presents a 3D design taxonomy (extraction, transmission, metrics).\\  \textbullet Discusses ML-based and knowledge base-based semantic modeling techniques.} & \CIRCLE & \CIRCLE & \CIRCLE & \DOWNcircle & \DOWNcircle \\ 

\cite{lu2023semantics} & \parbox[c]{23em}{\textbullet Provides an in-depth tutorial and survey on semantic information theory and reasoning-based communication. \\ \textbullet Introduces content/channel semantics and implicit/explicit reasoning.\\ \textbullet Connects theoretical metrics with AI-enabled designs.}& \CIRCLE & \CIRCLE & \CIRCLE & \CIRCLE & \DOWNcircle \\ 

\cite{chaccour2024less} & \parbox[c]{23em}{\textbullet Presents a reasoning-driven vision for AI-native semantic networks.\\ \textbullet Defines SemCom principles from broad theoretical level.\\ \textbullet Integrates causal reasoning and knowledge-centric design.} & \DOWNcircle & \CIRCLE & \DOWNcircle  & \DOWNcircle & \CIRCLE \\

\cite{zhang2024intellicise}& \parbox[c]{23em}{\textbullet Presents the concept of intellicise wireless networks from SemCom.\\ \textbullet Focuses on network-level integration.\\ \textbullet Reviews key technologies and enabling methods.} & \CIRCLE & \CIRCLE & \DOWNcircle  & \DOWNcircle &\Circle \\

\cite{getu2024survey} &  \parbox[c]{23em}{\textbullet Emphasizes goal-oriented SemCom frameworks and taxonomy. \\ \textbullet Reviews SemCom with broad theoretical and algorithmic coverage. \\ \textbullet Analyzes challenges and trends integrating 6G and SemCom.} & \CIRCLE & \CIRCLE & \DOWNcircle & \DOWNcircle & \DOWNcircle \\

\cite{getu2025semantic} & \parbox[c]{23em}{\textbullet Presents principles and models of SemCom for 6G. \\ \textbullet Focuses on principles, architecture, and conceptual modeling of SemCom. \\ \textbullet  Discusses semantic representation, system models, and open challenges.} & \CIRCLE & \CIRCLE & \DOWNcircle  & \DOWNcircle & \DOWNcircle \\

This survey & \parbox[c]{23em}{\textbullet Focuses on visual data-oriented SemCom, filling the gap left by general SemCom surveys.\\ \textbullet Categorizes diverse SemCom designs based on semantic quantization schemes and communication goals.\\ \textbullet Elaborates on the semantic quantization, encoder–decoder construction, and model training from a CV standpoint.\\ \textbullet Explores the structure and utilization of knowledge.} & \CIRCLE & \CIRCLE & \CIRCLE & \CIRCLE & \CIRCLE\\
\hline
  \end{tblr}

\smallskip
\textit{Notations:} \CIRCLE \, indicates fully included, \DOWNcircle \, means partially included, and \Circle \, means not included.
\end{table*}

\subsection{Contribution and Organization}
\begin{figure*}
    \centering
    \includegraphics[width=0.98\linewidth]{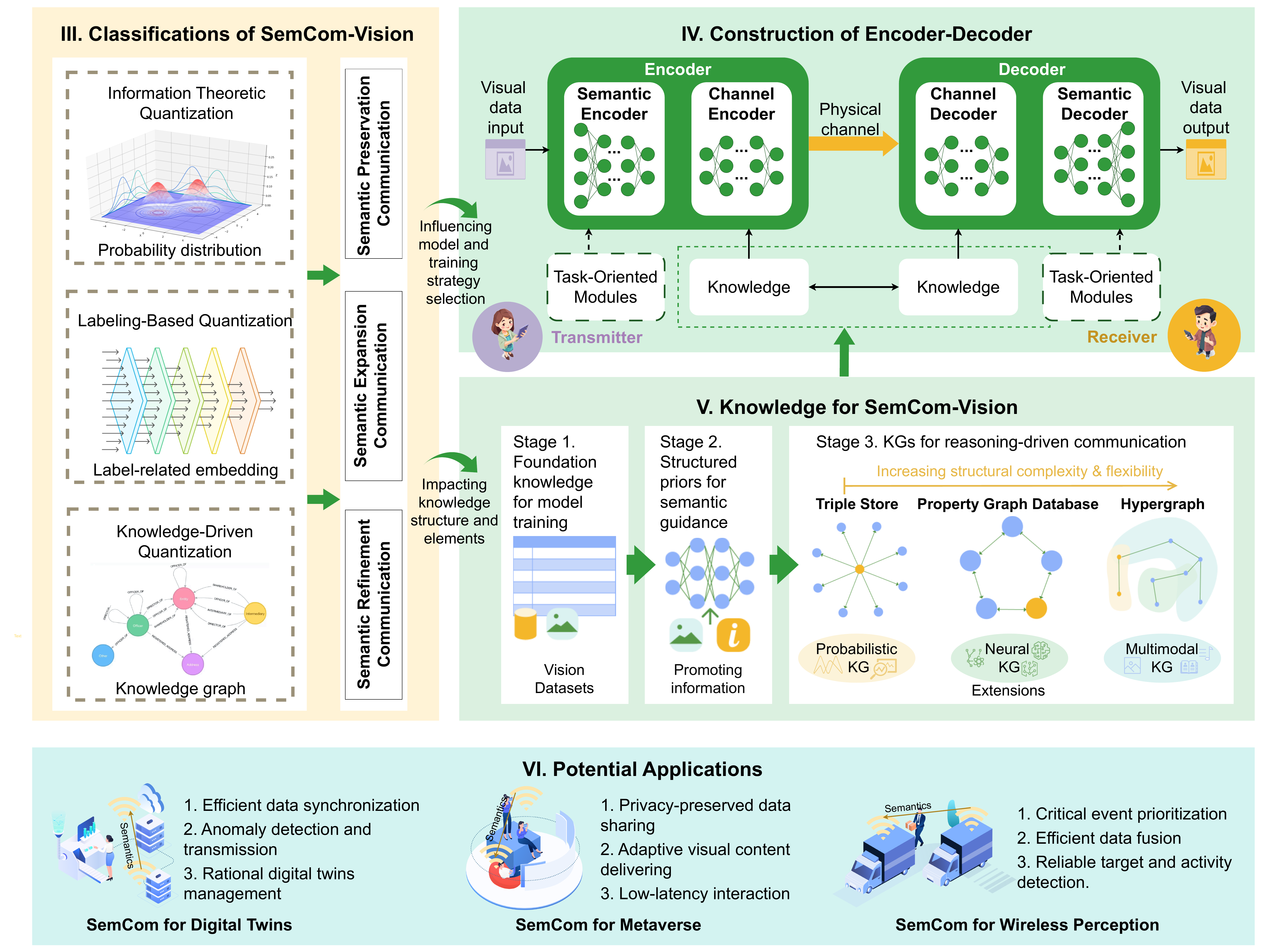}
    \caption{The construction of SemCom-Vision includes three major parts: 1) semantic quantization and SemCom classification, 2) encoder-decoder construction, 3) knowledge for SemCom-Vision. }
    \label{fig.1}
\end{figure*}

\minew{This survey provides a systematic review of SemCom for visual data transmission, offering several key contributions that distinguish it from existing literature:
\begin{itemize}
    \item This paper first explains the basics of ML in SemCom-Vision and introduces the holistic framework in SemCom-Vision transceiver design.  
    \item We introduce a novel classification method for SemCom-Vision approaches based on communication goals viewed through semantic quantization schemes, which provides a clear categorization of semantic preservation, expansion, and refinement communications, enabling researchers to better understand the landscape and identify research gaps.
    \item We provide an in-depth technical analysis of encoder-decoder construction specifically for SemCom-Vision, including different ML-based architectures and corresponding training algorithms, and cover both general theoretical foundations and task-specific considerations.
    \item We survey and examine knowledge structure and utilization essential for effective SemCom-Vision systems, including knowledge exploration stages and knowledge graph categorization.
    \item We explore emerging applications across multiple domains, including digital twins, metaverse, and wireless perception, reviewing how SemCom-Vision can be tailored to meet specific requirements and constraints of different application contexts.
\end{itemize}

The rest of this survey is organized as illustrated in Fig. \ref{fig.1}. Section \ref{sec:semcom_framework} introduces preliminary ML concepts and basic transceiver framework. Section \ref{sec:semantic_quantization} presents semantic quantization and the SemCom classification method for vision. Section \ref{sec:encoder_decoder} provides a detailed analysis of encoder-decoder construction strategies, including semantic preservation, expansion, and refinement techniques. Section \ref{sec:knowledge_management} discusses the knowledge structure and utilization essential for adaptive SemCom. Section \ref{sec:applications} explores practical applications across various domains, followed by the conclusion of this survey in Section \ref{sec:conclusion}.}

\section{Fundamentals of SemCom-Vision}\label{sec:semcom_framework} 

In this section, we introduce the basics of ML and the transceiver framework for the SemCom system, as depicted in Fig. \ref{fig.1}. 

\subsection{Preliminary: ML for SemCom-Vision }
ML is a transformative force that forms the backbone of SemCom-Vision  \cite{guo2024survey}. Particularly, NN-empowered ML excels in extracting, transmitting, and reconstructing semantics \cite{lan2021semantic}.
These NNs are generally built from some fundamental components, typically including fully connected layers, convolutional layers, activation functions, pooling layers, normalization layers, attention mechanisms, and residual and skip connections \cite{li2021survey}.

\textbf{Fully Connected Layer:}
A fully connected layer, also known as a dense layer, connects every neuron in the previous layer to every neuron in the subsequent layer \cite{ma2017equivalence}. The fully connected layer is mathematically represented as $ h = \sigma(Wx + b) $, where each hidden layer neuron $h$ computes a sum of its inputs $x$ with weight $W$ and adds a bias $b$ term before applying a non-linear activation function \cite{basha2020impact}. This layer is crucial for integrating information globally across the network, however, it often contains a large number of parameters, making large-scale SemCom models computationally complicated \cite{kocsis2022unreasonable}. 

\textbf{Convolutional Layer:}
The convolutional layer is designed to process grid-like structure data like images, thus being especially common in SemCom-Vision  \cite{long2015fully}. During convolution, a kernel slides over the input data to produce feature maps that capture essential local features like edges, textures, and shapes. The convolution operation can be mathematically described as $h_{i,j}^{(k)} = \sum_{m} \sum_{n} W_{m,n}^{(k)} \, x_{i-m, j-n} + b^{(k)}$, where $ W_{m,n}^{(k)} $ denotes the learnable kernel weights for the $k$th filter, $ b^{(k)} $ is the bias term, and $ x_{i,j} $ represents the input gird point \cite{o2015introduction}. The parameter sharing and local connectivity of the convolutional layer significantly reduce the number of parameters compared to fully connected layers, making it particularly efficient for vision-based processing tasks \cite{Wang2018understanding}.

\textbf{Activation Function:}
Activation functions introduce non-linearity into NNs, enabling them to model complex, non-linear relationships in the data \cite{nwankpa2018activation,tang2025finback,tang2025roby,tang2023pile}. Common activation functions include the rectified linear unit (ReLU), Sigmoid, hyperbolic tangent (Tanh), and Softmax. The ReLu, as one of the most popular choices in NN hidden layers, is defined as $ f(x) = \max(0, x) $. It passes positive inputs unchanged while setting negatives to zero, which produces sparse activations and helps prevent the vanishing gradient problem \cite{schmidt2020nonparametric}. The Sigmoid function $f(x) = \frac{1}{1 + e^{-x}}$ maps any input value to a range between 0 and 1. This characteristic makes it especially useful for binary classification tasks where outputs represent probabilities \cite{nwankpa2018activation}. The Tanh function is defined as $f(x) = \frac{e^{x} - e^{-x}}{e^{x} + e^{-x}}$. It maps the inputs to the range $[-1, 1]$, which helps in faster convergence during training by balancing positive and negative activations \cite{abdelouahab2017tanh}. The choice of activation function significantly affects the learning dynamics and convergence behavior of the network, as well as the semantic processing accuracy and efficiency \cite{mercioni2020most}.

\textbf{Pooling Layer:}
In vision-oriented SemCom, pooling layers are used to reduce the spatial dimensions of feature maps and to achieve translation invariance in the learned representations \cite{yu2014mixed}. Typically, pooling layers perform operations such as maximum or average pooling over small regions, thereby summarizing the most important features in each region. By reducing the spatial resolution, pooling layers decrease the computational burden and help prevent overfitting. This downsampling process is critical for extracting robust features that are less sensitive to small shifts or distortions in the input data \cite{liu2015treasure}.

\textbf{Normalization Layer:}
The normalization layer stabilizes and accelerates NN training by standardizing activations \cite{huang2023normalization}. For example, batch normalization normalizes outputs by adjusting for batch mean and variance, while variants like layer and instance normalization operate across different dimensions. This reduces internal covariate shift, enabling higher learning rates and faster convergence \cite{ba2016layer}.

\textbf{Attention Mechanism:}
Attention mechanisms allow a network to focus on the most relevant parts of the input by assigning higher weights to important features while reducing the influence of less significant ones \cite{niu2021review}. Self-attention, in particular, captures global dependencies by computing pairwise interactions between input elements \cite{vaswani2017attention }. By emphasizing informative regions, attention enhances both model interpretability and performance, especially in vision tasks requiring a strong understanding of contextual relationships \cite{dosovitskiy2020image}.

\textbf{Residual and Skip Connections:}
Residual and skip connections address the issue of vanishing gradients in deep networks by allowing the flow of information to bypass one or more layers. Residual connections add the input of a layer directly to its output, enabling the network to learn residual mappings rather than direct transformations \cite{targ2016resnet}. Similarly, skip connections pass high-resolution features from early layers to later layers, preserving detailed spatial information that is critical for tasks like semantic segmentation \cite{mao2016image}. Both types of connections facilitate efficient gradient propagation and help maintain the integrity of the learned representations throughout the network \cite{ronneberger2015u}.

\subsection{Basics of SemCom Transceiver Frameworks}
The SemCom transceiver, built upon the aforementioned fundamental NN components, serves as the backbone for semantic extraction, prioritization, transmission, and reconstruction \cite{qin2021semantic}. Unlike traditional transceivers that rely on source encoding-decoding and channel encoding-decoding built on deterministic rules or statistical models, the SemCom transceivers ensure more efficient and intelligent communication by integrating multiple ML-based modules, including the knowledge base, semantic encoder-decoder, channel encoder-decoder, and task-relevant evaluation and optimization module, as shown in Fig. \ref{fig.1}.

\textbf{Knowledge Base:}
Unlike traditional communication, which treats all transmitted data equally, SemCom relies on prior knowledge to minimize redundant transmission while ensuring accurate interpretation at the receiver \cite{zhao2022background}. The knowledge base is a shared repository of contextual, domain-specific, or general prior knowledge information used by both the transmitter and receiver. By leveraging this shared information, the knowledge base reduces redundancy, as parties reference common concepts instead of repeatedly transmitting foundational details. Moreover, it supports semantic reasoning, helping transceivers infer implicit information and adapt dynamically through ML or updates from interactions \cite{liu2025knowledge}.

\textbf{Semantic Encoder-Decoder:}
The semantic encoder-decoder plays a central role in extracting high-level semantic features from input data, compressing them for transmission, and reconstructing meaningful content at the receiver \cite{peng2025semantic, jiang2025lightweight,peng2025personalized}. Unlike traditional source encoding, which focuses on bit-level compression, semantic encoding prioritizes extracting goal-oriented or task-relevant semantics while discarding redundant or irrelevant binary bits \cite{xie2020lite}. Afterward, transmitted semantics can be reconstructed into desired content \cite{xu2023deep}. By optimizing feature extraction and reconstruction, semantic encode-decoders can avoid semantic loss or ambiguity and maintain high-efficiency and high-accuracy communication even under harsh channel conditions \cite{dong2022innovative}.

\textbf{Channel Encoder-Decoder:}
The channel encoder-decoder for SemCom ensures the robust transmission of semantics over noisy or unreliable channels by optimizing error correction and prioritizing key semantic features \cite{park2024joint}. Traditional channel coding techniques (e.g., Turbo, or Polar codes) focus on bit-level error correction, ensuring minimal loss of transmitted data without regard to data importance, contextual relevance, etc \cite{rowshan2024channel}. However, in SemCom, certain features may be more critical than others. Therefore, semantic-aware channel encoding adjusts redundancy and protection levels dynamically based on the priority of transmitted semantics \cite{wang2023improved}.

\textbf{Task-Oriented Modules:}
Task-oriented modules are optional supplemental modules in the SemCom transceiver, which are primarily designed to ensure that transmitted semantics optimize task performance \cite{ma2023task}.  Unlike conventional communication systems that merely assess transmission quality using naive signal metrics like bit error rate (BER), peak signal-to-noise ratio (PSNR), etc., SemCom evaluates information transmission comprehensively according to diverse tasks, where perception-aware metrics, e.g., classification accuracy, object detection precision, image perceptual quality, or aesthetics score are more important \cite{wang2024feature}. In this context, task-oriented modules could be exploited as evaluators to check the task-specific content quality requirement.
Moreover, task-oriented modules include components that can aid service requirement trade-off under varying channel conditions and resource availability \cite{ma2025power}. For example, ML-based resource allocation models to adjust computational and communication resource usage in semantic extraction,  transmission, and reconstruction, while considering the trade-off between service requirements like latency and content quality.

\section{Classifications of SemCom-Vision} \label{sec:semantic_quantization} 
The core principle of SemCom design is minimalism, i.e., minimizing the amount of data transmitted while preserving the essential semantics. However, with the differences in the explanation of essential semantics, the ML models used in SemCom transceivers are typically diverse. In this section, we elaborate on the semantic quantization schemes and three major types of SemCom with different definitions of essential semantics.

\begin{table*}[ht]
\centering
\caption{Comparison of Semantic Quantization Schemes for SemCom-Vision }
\label{table_2_quanti}
\begin{tblr}{
  colspec  = {
             Q[l,m,wd=2cm]  
             Q[l,m,wd=3.5cm]  
             Q[l,m,wd=3.5cm]  
             Q[l,m,wd=3.5cm]  
             Q[l,m,wd=3.5cm]  
             },
  rowsep   = 2pt,     
  stretch  = 1.2,     
  rowhead  = 1,       
  row{1}   = {halign=c}, 
  row{odd} = {bg=gray9}
}
\hline
  \textbf{Methods}
    & \textbf{Semantic Modeling}
    & \textbf{Quantization Metric}
    & \textbf{Advantages}
    & \textbf{Limitations} \\
\hline
  Information‐Theoretic Quantization
    & Semantics as probability distributions over abstract semantic features.
    & Semantic entropy, mutual information, etc.
    & Strong theoretical foundation, explains the limits of compression rate, channel capacity, etc.
    & Requires assuming distributions for semantics, complex to estimate distributions \\

  Labeling‐Based Quantization
    & Semantics are learned embeddings from vision-language models that are related to categorical labels, segmentation maps, etc.
    & Classification probabilities, attention weights, vector similarity, etc.
    & Easy to align with human-level tasks, intuitive, and practical in semantic extraction
    & Semantic granularity depends on label design or model training, model training requires large-scale labeled data. \\

  Knowledge‐Driven Quantization
    & Semantics as nodes and edges in a KG with structured relationships.
    & Graph‐based measures (e.g., cosine similarity of embeddings, path‐length metrics)
    & Captures more hierarchical and contextual information, enables reasoning and semantic inference
    & KG construction and maintenance overhead, sparse coverage in visual domains \\
\hline
\end{tblr}
\end{table*}

\subsection{Semantic Quantization for Vision}
Visual data is represented numerically as pixels, but its semantic interpretation relies on humans or intelligent agents who structurally annotate entities and their relationships using schemes such as object detection or region segmentation \cite{voulodimos2018deep, ye2023segmentation}. To move beyond pixel-level metrics, semantic quantization schemes are explored to measure the meaning embedded in visual data according to factors like relevance, importance, and robustness, as shown in Table \ref{table_2_quanti}.

\textbf{Information-Theoretic Quantization:} 
In essence, this type of method generally models semantics as probability distributions and then quantifies the total semantic content via entropy. Semantic entropy is the fundamental measure of semantics and underpins the development of SemCom systems \cite{islam2024deep}. Assume the set of semantics $\mathbf{s}$ is extracted from source input $\mathbf{x}$, and all the semantics are independent, i.e., $p(\mathbf{s})=\prod_{k=1}^{K}p({s}_{k})$. Moreover, for each semantic ${s}_{k}$ follows the probability distribution $p({s}_{k}|\mathbf{x})$ \cite{niu2024mathematical}. The semantic entropy $ H(\mathbf{s}) $ is then defined as
\begin{equation}
    H(\mathbf{s}) = -\sum_{k=1}^{K} p({s}_{k}) \log_{2} p(\mathbf{s}_{k}),
\end{equation}
where $K$ is the identified number of semantics. This semantic entropy quantifies the encoded semantics from uncertainty.

In practical SemCom systems, the extracted semantics $\mathbf{s}$ are transmitted over a channel to be reconstructed at the receiver as $\hat{\mathbf{s}}$. However, due to noise, interference, or compression, not all semantics can be preserved \cite{getu2024survey}. In this case,  semantic mutual information is exploited to quantify the effectiveness of semantic transmission, which is defined as 
\begin{equation}
    I(\mathbf{s};\hat{\mathbf{s}}) = H(\mathbf{s}) - H(\mathbf{s}|\hat{\mathbf{s}}), 
\end{equation}
where $H(\mathbf{s}|\hat{\mathbf{s}})$ measures how much uncertainty remains about the original semantics after observing the received semantics \cite{niu2024mathematical}. Therefore, $I(\mathbf{s};\hat{\mathbf{s}})$ captures the amount of semantics that are successfully preserved and transmitted. 
Building on the notions of semantic entropy and mutual information, some studies have proposed and formalized SemCom systems.
The authors in \cite{niu2024mathematical} provide a comprehensive exposition of semantic information measures grounded in semantic entropy. Additionally, the paper introduces the synonymous equipartition property and proves the semantic source coding theorem, the semantic channel coding theorem, and the semantic rate distortion coding theorem. It provides the foundation for entropy-based semantic quantization and offers key guidelines for the design of SemCom systems. In \cite{ma2023task}, the authors propose a task-oriented explainable SemCom framework grounded in semantic entropy. The framework derives the optimal input for rate-distortion-perception theory and explores the bounds of semantic channel capacity in the proposed SemCom system. The work in \cite{wu2024cddm} integrates a diffusion-based denoising model into SemCom, and utilizes conditional entropy to evaluate both the denoising capability of the proposed transceiver and the extent to which semantics are retained. Moreover, to quantifying semantics and further enhance semantic transmission, the information bottleneck (IB) principle is adopted as a theoretical framework for balancing compression and relevance in \cite{barbarossa2023semantic}, in which the goal is to minimize the mutual information $I(\mathbf{x}; \mathbf{s})$, which encourages compression, while maximizing $I(\mathbf{s}; \hat{\mathbf{x}})$, which ensures semantic fidelity. In \cite{wei2023federated}, the IB theory is employed to guide loss function design by formalizing a rate-distortion trade-off, effectively eliminating the redundancies of semantics while preserving task-relevant content. Additionally, the authors in \cite{xie2025advanced} propose a dynamic coding approach to make the tradeoff between compressing image $\mathbf{X}$ and preserving semantics $\hat{\mathbf{s}}$ for the image by adjusting the adaptive weights of IB-based SemCom, thus adjusting the extracted semantics in response to varying channel conditions. 

\textbf{Labeling-Based Quantization:} This type of method leverages ML models to map complex entities within visual data into low-dimensional spaces according to textual labels \cite{hua2025mathematical}. Trained on large-scale corpora with massive visual data-to-label pairs, these models are capable of capturing deep semantic relationships and contextual dependencies \cite{khan2022transformers}.  
For instance, the attention mechanism-based models dynamically allocate weights to different semantic features based on their contribution to the overall meaning \cite{dosovitskiy2020image}. This enables the models to focus on and enhance task-relevant semantics \cite{guo2022attention}. For example, in autonomous driving scenarios, higher attention weights can be assigned to the pedestrian regions compared to the background, effectively localizing critical targets \cite{han2022survey}. 
Moreover, some ML models like contrastive language-image pre-training (CLIP) \cite{radford2021learning} align the semantic spaces of text and images, mapping the textual description and its corresponding image to nearby embeddings. These embedding strip away redundant details like syntactic structures in text or pixel-level noise in images, preserving only the essential semantics \cite{shen2021much}. 

As one of the most general quantization methods, there are plenty of works that quantify semantics with labeling-based models. In \cite{pan2023image}, a Swin Transformer-based  \cite{liu2021swin} multi-scale semantic feature extractor is exploited in the SemCom encoder. This module leverages an attention mechanism to aggregate semantic features across multiple scales prior to transmission. In this framework, the semantics of pixels are mapped to category-level classifications of labels, enabling efficient, compact, and task-relevant semantic features. 
The work of \cite{huang2022toward} defines the semantic concept of image data that includes the category, spatial arrangement, and visual feature as the representation unit, and proposes a convolutional semantic encoder to extract semantic features.
In \cite{qian2023deep}, the authors employ an attention-based model to perform semantic segmentation across multiple categories. The categorization probabilities of the identified objects are used to quantify both the correctness and richness of the extracted semantics. Similarly, in \cite{li2025knowledge}, semantic features are obtained by leveraging the CLIP model, which encodes image inputs into text-related embedding vectors. Furthermore, \cite{nam2024language} introduces a SemCom framework enhanced by a large language model, where semantics are extracted in the form of text prompts and used to guide the generation of corresponding images in the receiver.

\textbf{Knowledge-Driven Quantization:} In this method, semantic features are mapped into nodes and relations in a knowledge graph (KG) \cite{chaccour2024less}. Knowledge-driven quantization differs from the aforementioned two methods, rather than treating semantics purely as probability distributions or as points in an abstract embedding space, it first maps each meaningful concept (objects, properties, etc.) onto nodes within a KG \cite{wang2017knowledge}. These nodes are connected by edges that capture various relationships, including hierarchical ($Animal\rightarrow cat \rightarrow Siamese \,\, cat$), part-whole ($cat \rightarrow cat \,\, furr$), or associative ($cat\rightarrow mouse$).
This mapping can leverage either information-theoretic principles (e.g., semantic entropy over graph-defined concept sets) or labeling-based scores (e.g., class probabilities assigned to graph nodes), or both simultaneously, as long as it anchors semantics in a web of human/intelligent agent-constructed knowledge \cite{zhou2023cognitive}.
After constructing the graph, semantic similarity is quantified by exploiting its topology and learned vector representations. Graph-embedding algorithms like TransE \cite{bordes2013translating}, where relations act as translations between node embeddings, or GraphSAGE \cite{hamilton2017inductive}, where the vector of each node is aggregated from its neighbors, transfer the discrete graph into a continuous space. Within this space, pairwise cosine similarities or other learned distance metrics reveal how semantically ``close" two semantics are, while path-based measures (such as shortest-path length weighted by edge importance) uncover the strength of more indirect connections \cite{ji2021survey}. While several studies have explored knowledge-driven quantization in text-oriented SemCom \cite{zhao2023semantic}, the knowledge-driven quantization for vision-oriented SemCom is still in its early stages and remains largely underexplored.

The authors in \cite{li2022domain} propose a knowledge extractor to process input images and output the knowledge semantics for a clinical diagnosis scenario. By linking text and image in KG, the proposed scheme can efficiently deliver medical images under harsh channel conditions. In \cite{xing2024representation}, the authors propose a multi-model semantic representation and fusion model based on a KG, in which the direct and reasoning correlation information is extracted and mapped into a two-layer semantic architecture to fully represent the semantics of each modality. 
The work of \cite{song2025uav} utilizes the auxiliary knowledge from the KG to quantify the semantics of unmanned aerial vehicle (UAV)-photoed images and establish associations with semantic features, thus enhancing accuracy in the task-oriented SemCom. 

Notably, rather than relying on a single quantization approach, the information-theoretic, labeling-based, and knowledge-driven methods can also be integrated together to leverage their respective strengths for more robust and rational semantic representation. For instance, a hybrid model may use attention-based models to extract label-relevant semantic features. Concurrently, the uncertainty and informativeness of the extracted semantics can be assessed using entropy or mutual information metrics. Then, refine or validate these features through alignment with structured knowledge from a KG. 

\subsection{SemCom-Vision Classification}
When viewed through the lens of semantic quantization, the three major types of SemCom-Vision naturally emerge according to the goals, as shown in Table \ref{Table_3_overview}.

\begin{itemize}
    \item \textbf{Semantic Preservation Communication (SPC)}: Transmit extracted semantics while mitigating channel-induced semantic loss or distortion and ultimately preserve pixel and semantic-level consistency between transmitted content and received content.
    \item \textbf{Semantic Expansion Communication (SEC)}: Extract semantics from input content and enrich transmitted semantics according to contextual or additional generative details, thus improving visual understanding and creativity of received content.
    \item \textbf{Semantic Refinement Communication (SRC)}: Extract and transmit only task-relevant semantics, reducing the richness and redundancy of the received content while improving alignment with task requirements and enhancing transmission efficiency. 
\end{itemize}

From the information‐theoretic perspective, SPC maximizes the retained semantic entropy $H(\mathbf{s})$ and mutual information $I(\mathbf{s};\hat{\mathbf{s}})$, ensuring that almost all original semantics survive transmission. SEC, by contrast, intentionally increases the entropy beyond that of the source by injecting new, contextually relevant information, effectively raising $H(\hat{\mathbf{s}})$ through generative augmentation. SRC leverages the IB principle to deliberately reduce $H(\mathbf{s})$, discarding redundant semantics and focusing on the subset that maximizes task‐relevant mutual information $I(\mathbf{s};\hat{\mathbf{s}}_{T})$, where $\hat{\mathbf{s}}_{T}$ denotes the task-specific outcome. 

Under the labeling-based quantization aspect, SPC maintains a stable set of labels for embeddings in the semantic extraction, transmission, and reconstruction, thus preserving the semantic features of visual data as closely as possible. In contrast, SEC enriches the visual data by injecting additional labels of semantics, whereas SRC selectively reduces the corresponding label number of semantics to keep bandwidth for more important information.

Moreover, from the perspective of knowledge-based quantization, SPC aims to preserve the fidelity of all semantic nodes and their corresponding vectors within the KG. In contrast, SRC selectively retains only the most task-relevant nodes and vectors, while SEC actively expands nodes and vectors of semantics from the KG to enrich the semantic features.

\subsection{Lessons Learned}
The diversity of semantic quantization approaches, spanning information-theoretic, labeling-based, and knowledge-driven paradigms, reveals that semantic quantification of visual data is inherently multifaceted, with a single scheme challenging to universally capture all aspects of visual semantics. Moreover, in SemCom-Vision, critical trade-offs exist between semantic fidelity, richness, and transmission efficiency. When pointing toward different communication goals, it is necessary to flexibly select appropriate quantization schemes and SemCom-Vision categories.

\section{Construction of SemCom Encoder-Decoder}  \label{sec:encoder_decoder} 
\begin{figure*}
    \centering
    \includegraphics[width=0.9\linewidth]{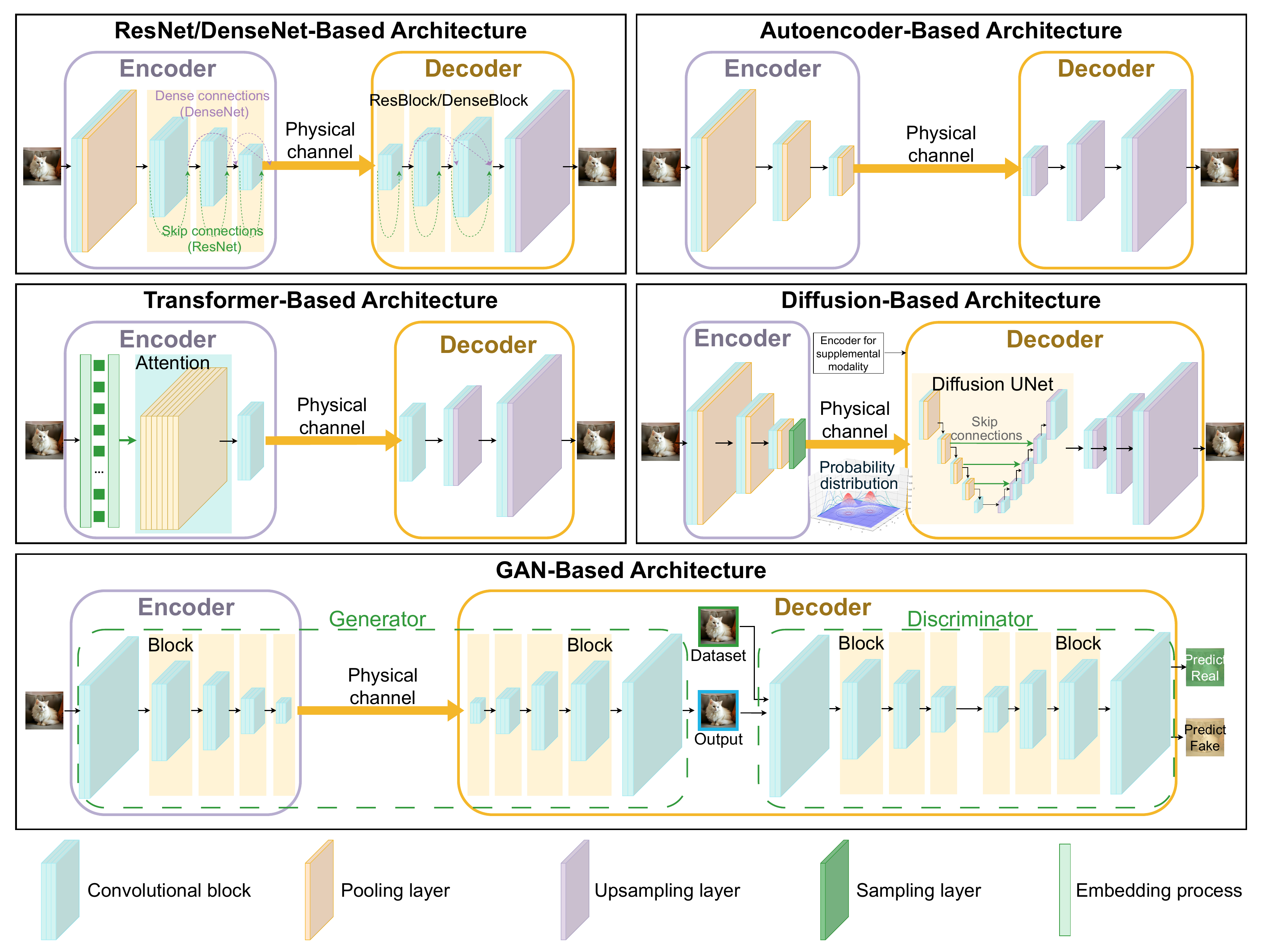}
    \caption{Typical ML model architectures for encoder-decoder design. }
    \label{fig.2}
\end{figure*}

The choice of encoder–decoder model in a SemCom-Vision system is fundamentally driven by the information transmission goal, whether it aims to preserve, expand, or refine the transmitted semantics, as summarized in Table \ref{Table_3_overview}. Although the core encoder–decoder paradigm remains consistent for SemComs of different communication goals, the specific ML models and NN architectures can be diversified, as shown in Fig. \ref{fig.2}. In this section, we separately elaborate on the ML-based encoder–decoder construction for the aforementioned three SemCom-Vision categories. 

\begin{table*}[!t]
\centering
\caption{Overview of SemComs and Corresponding Encoder-Decoder Models}
\label{Table_3_overview}
\begin{tblr}{
        colspec={Q[c,1cm] Q[l,2.8cm]Q[l,2.2cm]Q[l,3cm]Q[l,3cm]Q[l,1.5cm]Q[c,1.2cm]},
        cell{2}{1} = {r = 5}{c,m},
        cell{2}{2} = {r = 5}{l,m},
        cell{2}{3} = {r = 5}{l,m},
        cell{2}{7} = {r = 5}{l,m},
        cell{7}{1} = {r = 2}{c,m},
        cell{7}{2} = {r = 2}{l,m},
        cell{7}{3} = {r = 2}{l,m},
        cell{7}{7} = {r = 2}{l,m},
        cell{9}{1} = {r = 2}{c,m},
        cell{9}{2} = {r = 2}{l,m},
        cell{9}{3} = {r = 2}{l,m},
        cell{9}{7} = {r = 2}{l,m},
        row{odd} ={bg = gray9},
        column{1} ={bg = gray9}, 
        column{2} ={bg = white}, 
        column{3} ={bg = gray9},
        column{7} ={bg = gray9},
        row{1} ={bg = gray9},
        rowhead  = 1,
        row{1}   = {halign=c}
}
\hline
    \textbf{Category}  & \textbf{Goals and Features} & \textbf{Related CV Tasks/Applications}  & \textbf{Model Architectures and ML Models} & \textbf{Functions of Models}  &\textbf{SemCom References} &\textbf{Loss Functions}\\
    \hline

    Semantic Preservation Communication (SPC) 
    & Transmit extracted semantics while mitigating channel-induced semantic loss or distortion, and ultimately preserve pixel- and semantic-level consistency.
    & Image reconstruction / super-resolution, image classification, semantic segmentation, object detection.   
    & ResNet-based: ResNet \cite{he2016deep}, WRN \cite{zagoruyko2016wide}, SE-ResNet \cite{hu2018squeeze}. \newline
    DenseNet-based: DenseNet \cite{huang2017densely}, CondenseNet \cite{huang2018condensenet}. 
    & Hierarchical feature extraction, residual and dense connectivity for robust gradients
    & \cite{zhang2023deepma},\cite{tang2024contrastive},\cite{zhang2024compression}
    & $\mathcal{L}_{1}$, $\mathcal{L}_{2}$, $\mathcal{L}_{3}$, $\mathcal{L}_{4}$, $\mathcal{L}_{5}$, $\mathcal{L}_{6}$. \\
    &
    &
    & Autoencoder-based: VAE \cite{kingma2013auto}, VQ-VAE \cite{van2017neural}, VQ-VAE-2 \cite{razavi2019generating}, CVAE \cite{harvey2021conditional}, DAE \cite{bengio2013generalized}
    & Dimensionality reduction, latent semantic extraction, and reconstruction
    & \cite{zhang2025progressive}, \cite{si2024post}
    & \\

    &
    &
    & Diffusion-based: DDPM \cite{ho2020denoising}, DDIM \cite{song2020denoising}, SGM \cite{song2020score}
    & Stochastic denoising, robust semantic completion
    & \cite{li2024goal}, \cite{wu2024cddm}, \cite{guo2025diffusion}
    & \\
    &
    &
    & Transformer-based: ViT \cite{dosovitskiy2020image}, Swin Transformer \cite{liu2021swin} 
    & Global context modeling, attention-based semantic aggregation
    & \cite{peng2024robust}, \cite{zhang2023predictive}, \cite{zhao2024lamosc}
    & \\
    &
    &
    & GAN-based: GAN \cite{goodfellow2020generative}, DCGAN \cite{radford2015unsupervised}, WGAN \cite{arjovsky2017wasserstein}
    & Adversarial refinement, high-fidelity detail restoration
    & \cite{xin2024deep}
    & \\
    \hline

    Semantic Expansion Communication (SEC)
    & Extract semantics from input content and enrich transmitted semantics via contextual or generative augmentation, improving the visual understanding and creativity. 
    & Text-to-image generation, conditional image generation, image-to-image translation  
    & Diffusion-based: LDM \cite{rombach2021highresolution}, DALL-E \cite{ramesh2022hierarchical}, Imagen \cite{saharia2022photorealistic}, ControlNet \cite{zhang2023adding}
    & Transform noise into structured data, multimodal alignment, multimodal information fusion
    & \cite{zhang2024diffusion}, \cite{liang2024image}, \cite{hosonuma2024image}, \cite{fan2024semantic}
    &$\mathcal{L}_{4}$, $\mathcal{L}_{6}$, $\mathcal{L}_{7}$. \\
    &
    &
    & GAN-based: CGAN \cite{mirza2014conditional}, Pix2Pix \cite{isola2017image}, CycleGAN \cite{zhu2017unpaired}, StyleGAN \cite{karras2019style}, SRGAN \cite{zhang2019self}
    & Generate realistic and diverse data, unsupervised feature learning, stabilize training, and content regularization
    & \cite{lokumarambage2023wireless} 
    & \\
    \hline

    Semantic Refinement Communication (SRC)
    & Transmit only task-relevant semantics, reducing redundancy while maximizing alignment with downstream tasks for improved efficiency.
    & Semantic segmentation, salient object detection, visual question answering, region-of-interest selection
    & Transformer-based: MAE \cite{he2022masked}, PVT \cite{wang2021pyramid}, CVT \cite{wu2021cvt}, LXMERT \cite{tan2019lxmert}, VisualBERT \cite{li2019visualbert} 
    & Masked pretraining for targeted features, vision-language reasoning, and semantic importance prediction 
    & \cite{kang2022personalized, pan2023image, xie2021task, xie2022task}
    & $\mathcal{L}_{8}$, $\mathcal{L}_{9}$, $\mathcal{L}_{10}$, $\mathcal{L}_{11}$, $\mathcal{L}_{12}$. \\
    &
    &
    & ResNet-based: CBAM-ResNet \cite{woo2018cbam} \newline
      DenseNet-based: DenseNet \cite{huang2017densely}, CondenseNet \cite{huang2018condensenet}
    & Attention-driven channel-spatial focus, compact deep representations
    & \cite{qian2023deep,jiang2025high}
    & \\
    \hline
\end{tblr}
\end{table*}

\subsection{Encoder-Decoder for Semantic Preservation}  

The goal of SPC is to transmit information while ensuring high fidelity between the original and received content, both at the pixel and semantic levels. To achieve semantic consistency, the encoder-decoder architecture in SPC must be capable of effectively extracting comprehensive semantics from the input while being robust to transmission-induced distortions such as channel noise. Furthermore, it is also important for the encoder-decoder to maintain sensitivity to visual features, including color accuracy, structural integrity, and resolution, to preserve pixel-level fidelity.

\textbf{Model Architectures:}
To achieve the goal of semantic preservation, SPC is closely aligned with traditional CV tasks such as image reconstruction, super-resolution, image classification, semantic segmentation, and object detection, all of which require rich semantic understanding and detailed visual features \cite{chen2020survey}. Consequently, the five model architectures as shown in Fig. \ref{fig.2} commonly employed in these tasks are frequently adapted and integrated into the encoder-decoder design of SPC systems. 

\textit{ResNet/DenseNet-based architectures} are widely used backbone architectures in ML and have also been adopted in SPC systems for their powerful feature extraction capabilities. These convolutional NNs are architecturally similar and effective at capturing hierarchical visual features by exploiting residual or skip connections to preserve detailed semantics. As listed in Table \ref{Table_3_overview}, in SPC, models such as ResNet \cite{he2016deep}, WRN \cite{zagoruyko2016wide},  SE-ResNet \cite{hu2018squeeze}, DenseNet \cite{huang2017densely}, and CondenseNet \cite{huang2018condensenet}, are often employed as encoders. However, since their outputs are typically high-dimensional feature vectors intended for classification or recognition tasks, they are not inherently designed for compression or robust transmission. To be effective in SPC, these architectures are generally jointly trained with corresponding decoders under joint source-channel coding (JSCC) pipelines. \minew{Generally, these decoders are also composed of multiple ResBlocks \cite{he2016deep} or DenseBlocks \cite{huang2017densely},} as shown in Fig. \ref{fig.2}.
In \cite{zhang2023deepma}, the authors propose a ResNet-based encoder that transforms input data into semantics as symbol vectors. Additionally, it jointly trains the encoders and corresponding decoders to minimize the average distortion between the original and reconstructed data. The authors of \cite{tang2024contrastive} employ ResNet in both the encoder and decoder to ensure semantic preservation in both extraction and reconstruction.  Meanwhile, the work of \cite{zhang2024compression} presents a ResNet-based image sequence (video) compressed sensing SemCom, where the encoder and decoder are jointly trained to extract and transmit the semantics of video frames at a flexible compression rate, and further accurately reconstruct as pixel-fidelity video.

\textit{Autoencoder-based architecture} serves as one of the most important architectures for SPC design. Autoencoder-based models, including VAE \cite{kingma2013auto}, VQ-VAE \cite{van2017neural}, VQ-VAE-2 \cite{razavi2019generating}, CVAE \cite{harvey2021conditional}, and DAE \cite{bengio2013generalized}, learn compact latent representations that capture the most salient visual and semantic features of an image. During transmission, only those latent space semantics need to be sent, greatly reducing bandwidth requirements. At the receiver, the decoder reconstructs the image from the latent semantics, with the learned bottleneck ensuring that Channel noise-caused distortions are filtered out, thereby preserving pure semantics and ensuring pixel fidelity. 
In \cite{zhang2025progressive}, the authors propose a progressive image transmission framework that adaptively maps each semantic to a variable number of channel symbols. This mapping is guided by the entropy-aware priors learned through a hierarchical VAE, facilitating efficient allocation of communication resources. The authors in \cite{si2024post} propose a post-deployment fine-tunable SemCom system to enhance the adaptability of SemCom models to unseen datasets. Their framework is built upon a two-layer hierarchical VQ-VAE-2 autoencoder, which enables the extraction of robust and accurate semantics for further fine-tuning and faithful reconstruction. 

\textit{Diffusion-based architecture}, as a supplemental design in SPC, is often integrated with autoencoders, particularly VAEs. As shown in Fig. \ref{fig.2}, the encoder extracts semantic features from the image as a Gaussian-like probabilistic distribution, and then samples semantics from the distribution. \minew{More importantly, the diffusion UNet \cite{ronneberger2015u} is used to denoise semantic noise,} especially that caused by channel noise, thus ensuring that delivered semantics are consistent with the original input. After that, the denoised semantics can be reconstructed as images.
For semantic preservation, models like DDPM \cite{ho2020denoising} and DDIM \cite{song2020denoising} exhibit a strong denoising capability in the latent space, thus refining semantics and enhancing robustness to channel noise. In \cite{li2024goal}, the authors propose a goal-oriented SemCom framework that leverages a VAE-assisted diffusion model to iteratively purify and reconstruct semantic features. This hybrid approach enables more accurate and robust image reconstruction under varying channel conditions. By combining the compact representation learning of VAEs with the generative refinement power of diffusion models, their system demonstrates improved semantic preservation and pixel-level consistency. In \cite{wu2024cddm}, a channel denoising diffusion model is proposed as a physical layer module after the channel equalization to learn the distribution of the channel input signal, and then utilizes the learned knowledge to remove the channel-caused noise. Additionally, a diffusion-based transceiver is proposed in \cite{guo2025diffusion} to reduce channel-caused semantic noise by the denoising process, and ensure both pixel and semantic level consistency between images. 

\textit{Attention-based architecture} has also emerged as a powerful supplemental component in SPC, especially for capturing long-range dependencies and enhancing semantic feature alignment. Attention mechanisms, particularly those used in Transformer-based models, dispart and process images as different embeddings, as shown in Fig. \ref{fig.2}. Then, the encoder analyses and focuses on the most informative regions of the input, facilitating more accurate semantic extraction. The transmitted semantics are then reconstructed by a convolution block-based decoder, which is jointly trained with the encoder. \minew{By accurately retaining essential semantic features, attention modules improve robustness to semantic distortion and misalignment.} 
The work of \cite{peng2024robust} exploits an attention architecture in the encoder to identify the semantic importance of different embeddings, thus directing the focus toward the most relevant and semantically meaningful areas in semantic transmission. In \cite{zhang2023predictive}, the authors utilize the attention model to guide the semantic extraction and transmission in the ResNet-based encoder-decoder, which helps improve the transmission efficiency and ensure accurate semantic delivery.
In \cite{zhao2024lamosc}, a vision transformer (ViT) model is exploited to extract the image as the embedding, while an attention-based model is employed to derive the corresponding textual prompt. By fusing these accurately aggregated multimodal semantics, the system enhances the expressiveness and robustness of the transmitted representation, enabling more accurate image reconstruction and improved task-relevant semantic understanding.

\minew{\textit{Generative adversarial network (GAN)-based architecture} is generally used as a supplemental module to enhance the semantic and pixel consistency of input and output.} \minew{In SPC, pixel consistency is essential, however, traditional pixel-level distortion metrics, such as mean squared error (MSE) or PSNR, may fail to capture differences of small yet semantically critical regions.} GANs address this limitation by introducing a discriminator that guides the generator toward producing image details that are not only visually similar but also semantically consistent with the source. As shown in Fig. \ref{fig.2}, the GAN generator is typically integrated into the encoder-decoder backbones with other model architectures, where the discriminator serves as a module that refines image reconstructions. There are a few works that exploit GAN-based architecture in supporting SPC. 
For example, in \cite{xin2024deep}, the authors propose a GAN-based SemCom system for image transmission, where the generator is modified and separated into the encoder and decoder to extract semantics and reconstruct images, while the discriminator is employed to assess semantic similarity between the original and reconstructed images. 

It is important to note that these model architectures are not necessarily used in isolation. SPC systems can flexibly integrate multiple architectural components based on task requirements, channel conditions, image modalities, etc. This hybrid design strategy allows systems to harness the complementary strengths of different models. The convolutional backbones like ResNet and DenseNet provide powerful feature extraction, while autoencoders enable effective compression and reconstruction. Supporting modules such as attention mechanisms enhance relevance weighting, diffusion models offer robust denoising, and GANs improve perceptual quality. Ultimately, the choice and integration of architectures should align with the core objective of semantic preservation, ensuring both pixel-level fidelity and semantic consistency in the delivered images.

\textbf{Model Training:}  
Training the SPC encoder-decoder requires carefully designed loss functions that guide the network to maintain pixel fidelity and semantic consistency, while optimizing for transmission efficiency. 

Pixel-aware loss function: As common functions in traditional communication, they also find their place in SPC, because if every pixel is the same, the two figures are semantically consistent. To ensure that the reconstructed image remains numerically similar to the original, the following training losses are exploited in some existing works. 

The MSE loss function minimizes the squared differences between pixel values in the original and reconstructed images,
\begin{equation}
\mathcal{L}_{1} = \frac{1}{N} \sum_{i=1}^{N} (\mathbf{x}_i - \hat{\mathbf{x}}_i)^2, 
\label{L_1}
\end{equation}
where $ \mathbf{x}_i $ represents the pixel value in the original image, $ \hat{\mathbf{x}}_i $ is the corresponding pixel in the reconstructed image, and $ N $ is the total number of pixels \cite{wu2024cddm, zhang2024compression}. Although MSE ensures numerical accuracy, it often results in excessive smoothness and loss of fine details.

In this context, Charbonnier loss is a robust alternative to MSE that is less sensitive to outliers:
\begin{equation}
\mathcal{L}_{2} = \frac{1}{N} \sum_{i=1}^{N} \sqrt{(\mathbf{x}_i - \hat{\mathbf{x}}_i)^2 + \mathbf{\epsilon}^2},
\end{equation}
where $ \mathbf{\epsilon} $ is a small constant added for numerical stability \cite{cao2025task}. This loss reduces the influence of large deviations, preserving sharp edges and textures more effectively than MSE.

Moreover, the structural similarity index (SSIM) loss focuses on preserving high-level visual features that are important for human perception, like luminance, contrast, and structure, thus reducing perceptual distortions and preserving image details.
\begin{equation}
\begin{aligned}
    &\mathcal{L}_{3} = 1 - \mathit{SSIM}(\mathbf{x}, \hat{\mathbf{x}}), \\
    &\mathit{SSIM}(\mathbf{x}, \hat{\mathbf{x}}) = \left( \frac{2\mu \hat{\mu} + C_1}{\mu^2 + \hat{\mu}^2 + C_1} \right) \cdot \left( \frac{2\tilde{\sigma} + C_2}{\sigma^2 + \hat{\sigma}^2 + C_2} \right),
\end{aligned}
\end{equation}
where $\mu$ and $\hat{\mu}$ are the means of $\mathbf{x}$ and $\hat{\mathbf{x}}$, while $\sigma^2$ and $\hat{\sigma}^2$ denote the variances \cite{huang2022toward}. Moreover, $\tilde{\sigma}$ is the covariance between $x$ and $\hat{x}$, $C_{1}$ and $C_{2}$ are stability constants. 

{Semantic-pixel-aware loss function:} While pixel-level loss functions focus on ensuring numerical fidelity between the original and reconstructed images, they may not fully capture the semantic or perceptual essence of the content in SPC. To address this gap, semantic-pixel-aware loss functions have been introduced, leveraging deep feature representations to better align reconstructions with high-level semantics while preserving high pixel consistency. 

One of the typical semantic-pixel-aware losses is the MSE and Kullback–Leibler (KL)-divergence/variational loss, which is generally used in VAE-based model training 
\begin{equation}
\begin{aligned}
    \mathcal{L}_{4} = &\frac{1}{N}\sum_{i=0}||\mathbf{x}_i - \hat{\mathbf{x}}_{i}||^2 - \\ 
    &\frac{1}{2}\sum_{i=1}^{N}\sum_{j=1}^{K} 
    \left(1+log\left(\sigma_{i,j}^{2}\right)-\mu_{i,j}^{2}-\sigma_{i,j}^{2}\right).
\end{aligned}
\end{equation}
In this loss function, the first reconstruction term component, 
$\frac{1}{N}\sum_{i=0}||\mathbf{x}_i - \hat{\mathbf{x}}_{i}||^2$, generally the MSE, ensures the reconstructed output $\hat{\mathbf{x}}$ is close to the input $\mathbf{x}$. The second component is the KL divergence, calculated between the learned latent distribution of semantics and a standard Gaussian prior, where $\mu$ and $\sigma$ denote the mean and standard deviation of the encoded semantics \cite{guo2025diffusion}. This regularization term enhances the consistency of semantics from transmitter to receiver. 

\minew{In addition, learned perceptual image patch similarity (LPIPS) loss is another common function used along with other pixel-aware loss functions to measure both semantic and pixel differences}
\begin{equation}
\mathcal{L}_{5} = ||\phi(\mathbf{x}) - \phi(\hat{\mathbf{x}})||^2 + \mathcal{L}_{P} =||\mathbf{s} - \hat{\mathbf{s}}||^2+\mathcal{L}_{P}, 
\end{equation}
where the left part is LPIPS loss and $\mathcal{L}_{P}$ is pixel-aware loss function, such as MSE \cite{peng2024robust}. Additionally, $\phi$ is the ML NN (e.g., ResNet) for semantic extraction. This loss function effectively captures high-level perceptual discrepancies, making it useful for tasks where fine textures and high-frequency details need to be preserved.

Moreover, as a specialized loss function for GAN-based training, adversarial loss uses a discriminator to distinguish real images from generated ones, where the loss is given by
\begin{equation}
\mathcal{L}_{6} = \mathbb{E}\left(\log D\left(\mathbf{x}\right)\right) + \mathbb{E}\left(\log \left(1 - D\left(G\left(\mathbf{s}\right)\right)\right)\right), 
\end{equation}
where $ D(\mathbf{x}) $ is the discriminator's probability estimate that $ \mathbf{x} $ is a real sample, and $ G\left(\mathbf{s}\right) $ is the generator's output \cite{xin2024deep}. The generator aims to produce images that are indistinguishable from real samples, helping to reduce pixel blurring and improve perceptual quality.

\subsection{Encoder-Decoder for Semantic Expansion}

The objective of the SEC is to deliver content that not only aligns semantically with the source but also enriches the receiver's understanding, imagination, or task-specific performance. Compared to SPC, which emphasizes strict both semantic and pixel fidelity, SEC prioritizes semantic depth, contextual diversity, and generative richness. Therefore, in the SEC design, the encoder is tasked with extracting the essential high-level semantics from the input, while the decoder goes beyond simple reconstruction by generating enhanced outputs. 

\textbf{Model Architectures:}
To achieve semantic expansion, SEC leverages generative and context-aware model architectures closely related to generative AI tasks, such as conditional image generation, image-to-image translation \cite{zhao2024enhancing}. The diffusion-based and GAN-based architectures provide the capacity to decode extracted semantics into rich, creative, and contextually aligned outputs, enabling SEC to enhance the semantic richness and pixel-perceptual quality of delivered content.

\textit{Diffusion-based architecture} plays a central role in SEC due to its exceptional ability to generate high-quality and contextually rich content from noisy or incomplete latent space representations, i.e., semantics \cite{rombach2021highresolution}. Typically, the core components in this architecture are a semantic encoder, a diffusion UNet module, and a VAE decoder, as depicted in Fig. \ref{fig.2}. In the SEC framework, the encoder extracts latent semantics and corresponding conditional prompts from input images and additional modalities (such as textual descriptions and depth maps), respectively. Then, the prompts serve as contextual guides that enrich the generative process. During transmission, the semantics may be corrupted by channel noise or intentionally replaced with sampled noise for generative reconstruction \cite{liang2024image}. The diffusion UNet is then employed at the receiver to iteratively denoise these inputs, using conditional prompts to guide the reverse diffusion process. This enables the decoder to reconstruct images that are enriched with new semantic details derived from the provided context, thereby enhancing the visual expressiveness and creativity of the output. As listed in Table \ref{Table_3_overview}, models like LDM \cite{rombach2021highresolution}, DALL·E \cite{ramesh2022hierarchical}, Imagen \cite{saharia2022photorealistic}, and ControlNet \cite{zhang2023adding} are particularly suitable for SEC, as they enable structured content synthesis through fine-grained control over generation. 
In \cite{zhang2024diffusion}, the diffusion-based encoder-decoder is utilized to extract semantics from dual-fisheye images and then merge the semantics to stitch, expand, and generate panoramic images.
The work in \cite{liang2024image} explores semantic expansion by integrating latent diffusion with task-relevant prompts, thereby enriching image-to-text and text-to-image generation in transmission scenarios. \minew{Similarly, the work in \cite{hosonuma2024image} employs a VAE-assisted diffusion decoder that uses multimodal semantics extracted by the encoder to iteratively refine and expand visual information under channel constraints, supporting detailed and plausible scene reconstructions even from limited or abstract inputs.} In \cite{fan2024semantic}, the input images are decomposed into their natural language description, texture, and color semantic features at the transmitter. After being transmitted over the wireless channel, at the receiver, a diffusion UNet-based visual generation model is utilized to restore the image through received multimodal features. 

\textit{GAN-based architecture} also contributes significantly to SEC, especially in scenarios where realism and perceptual quality are paramount. Unlike in SPC, where the discriminator serves mainly to refine visual details or reduce distortion in the generator reconstructed outputs, in SEC, the generator is conditioned on one or more modalities to generate another, enabling cross-modal synthesis and semantic expansion. Moreover, the discriminator actively enhances, complements, or creatively reinterprets the transmitted semantics based on contextual prompts or latent priors. 
For example, in \cite{lokumarambage2023wireless}, a GAN-based SEC system generates artistically styled images from segmentation maps by incorporating both image features and global knowledge. The discriminator in this setup acts not just as a judge of realism but also as a semantic guide, ensuring that the generated output preserves and expands upon the original semantics. 

\textbf{Model Training:}
Training an SEC encoder-decoder system requires a carefully constructed set of loss functions that encourage semantic consistency and generative quality. These losses often combine objectives from unsupervised generative modeling and perceptual optimization.

Denoising loss as the core training function is used in diffusion-based SEC models, where the model is trained to reverse the noise process applied to a sample. These losses are typically formulated as:
\begin{equation}
\mathcal{L}_{7} = \mathbb{E}_{\mathbf{s}, \boldsymbol{\epsilon} \sim \mathcal{N}(0, 1)}\left[ \| \boldsymbol{\epsilon} - \boldsymbol{\epsilon}_\theta(\mathbf{s}_t, \check{\mathbf{s}}_t,t) \|^2 \right],
\end{equation}
where $\mathbf{s}_t$ is the semantics or noise of time $t$, $\check{\mathbf{s}}_t$ is the semantics extracted from supplemental modalities \cite{zhang2024diffusion}. 
This loss ensures that the model learns to progressively reconstruct meaningful content from corrupted or compressed representations.
It is noted that the denoising loss is generally used with VAE training loss, i.e., MSE-KL divergence loss $\mathcal{L}_{4}$. Different from SPC, which uses MSE-KL divergence loss mainly for faithful reconstruction, SEC exploits this loss to support creative, controllable generation while still grounding outputs semantically.

\minew{In SEC, adversarial loss $\mathcal{L}_{6}$ also plays a central role in guiding the generation of high-fidelity, visually realistic content that extends beyond the originally transmitted semantics. }


\subsection{Encoder-Decoder for Semantic Refinement}
The goal of SRC is to transmit only those semantic features that are strictly necessary for a downstream task, discarding redundant or irrelevant information to maximize efficiency. Unlike SPC, which seeks both pixel and semantic level fidelity, or SEC, which enriches and expands semantics, SRC focuses on precision over completeness. It must identify, encode, and transmit a minimal set of task-relevant features while ensuring those features retain their semantics under channel distortions.

\textbf{Model Architectures:}
\minew{SRC builds on architectures that are designed to disentangle and select the most relevant features and encode them in compact semantics. It is closely connected to CV tasks such as semantic segmentation, salient object detection, visual question answering, and region-of-interest selection, where a common objective is to remove redundant visual information while preserving the essential semantics.}

\textit{Transformer-based architecture:}
Transformers, particularly those based on MAE \cite{he2022masked}, PVT \cite{wang2021pyramid}, CVT \cite{wu2021cvt}, LXMERT \cite{tan2019lxmert}, and VisualBERT \cite{li2019visualbert}, are well-suited to SRC due to their strong capability in capturing global contextual dependencies and supporting selective attention mechanisms. In SRC, Transformer-based encoders attend to task-relevant regions or object queries, enabling efficient extraction and representation of high-value semantics. These models can adaptively filter and transmit only essential tokens or embeddings based on the task requirements. For instance, \cite{kang2022personalized} proposes a lightweight selective SemCom system for unmanned aerial vehicles where a Transformer encoder filters image patches based on tasks, reducing transmission load without compromising performance. The authors of \cite{pan2023image} propose a Transformer-based transceiver framework for the Internet of vehicles, which accurately extracts and transmits essential semantics related to vehicle control tasks, thus reducing the bandwidth resource storage and ensuring controlling task performance. In \cite{xie2021task, xie2022task}, similar Transformer-based architectures are used to represent images with minimal tokens optimized for robust semantic extraction and transmission over noisy channels.

\textit{ResNet/DenseNet-based architecture:}
ResNet/DenseNet and its variants are frequently used as backbones in SRC, especially when coupled with modules for region proposal or semantic filtering. In this context, the NN is adapted to focus on the most discriminative and task-relevant visual features. For instance, ResNet/DenseNet encoders can isolate semantic features aligned with specific tasks (e.g., object classification and anomaly detection). These features are then compressed and transmitted as prioritized semantics, discarding task-irrelevant information. It is noted that, in SRC, ResNet/DenseNet-based architecture is usually used with Transformer since it is powerful in semantic segmentation.
In \cite{qian2023deep}, a ResNet encoder is integrated with a semantic attention mask to isolate and encode only salient visual cues relevant to classification under constrained bandwidth conditions. Similarly, \cite{jiang2025high} utilizes a ResNet-based encoder with pruning and task-relevant heads to extract refined semantic tokens, and achieves high accuracy in remote sensing image scene classification tasks.

\textbf{Model Training:}
Training encoder-decoder models for SRC requires specialized loss functions that emphasize relevance, compactness, and robustness. These losses are designed to ensure that the transmitted features are both semantically meaningful and minimal, without carrying unnecessary details.

{IB loss:}
To explicitly encourage semantic compression, the IB principle is applied as a supplemental loss for other loss functions in many SRC frameworks. The loss is formulated as
\begin{equation}
\mathcal{L}_{8} = \mathcal{L}_{S} + \beta \cdot I(\mathbf{s};\mathbf{x}),
\end{equation}
where $\mathcal{L}_{S}$ is the original loss function used to measure semantic consistency, such as the LPIPS loss, and $I(\mathbf{s};\mathbf{x})$ is the mutual information between the encoded semantics $\mathbf{s}$ and input $\mathbf{x}$, and $\beta$ controls the trade-off. Minimizing $I(\mathbf{s};\mathbf{x})$ encourages the encoder to discard irrelevant input information while retaining task-relevant semantics \cite{wei2023federated}.

{Selective semantic loss:}
In Transformer-based SRC systems, selective semantic loss functions are used to enforce consistency only over semantically selected regions or tokens. For instance, a mask-aware reconstruction loss may only apply reconstruction penalties to selected semantic regions
\begin{equation}
\mathcal{L}_{9} = \frac{1}{|\Omega|} \sum_{i \in \Omega} | \mathbf{x}_i - \hat{\mathbf{x}}_i |^2,
\end{equation}
where $\Omega$ denotes the set of task-relevant pixel or token indices determined by attention weights or semantic masks \cite{qian2023deep}. This loss allows systems to focus on precise reconstruction where it matters and ignore irrelevant areas.

{KL-divergence and sparsity regularization loss:}
To maintain minimalism in representation, KL-divergence and sparsity-inducing regularizers are incorporated. These encourage the encoder to use fewer binary bits while still encoding sufficient semantic content for the task. For instance,
\begin{equation}
\mathcal{L}_{10} = \lambda \sum_{j=1}^{K} |\mathbf{s}_j| + \gamma D_{KL}\left(q\left(\mathbf{s}|\mathbf{x}\right) || q\left(\mathbf{s}\right)\right),
\end{equation}
where $q(\mathbf{s}|\mathbf{x}) $ is the learned semantic distribution and $q(\mathbf{s})$ is a prior (e.g., sparse or Gaussian) \cite{xie2024robust}. Additionally, $\lambda$ and $\gamma$ are factors to balance the first term for promoting feature sparsity and the second term for regularizing semantics for efficient encoding.

\minew{{Cosine similarity loss:} 
Some SRC systems focus only on the delivery of basic visual element categories, rather than transmitting all high-dimensional visual features. In such cases, images can first be mapped to corresponding textual vectors or semantic embedding vectors, and trained using cosine similarity loss
\begin{equation}
\mathcal{L}_{11} = 1 - \frac{s\cdot\hat{s}}{\|s\| \|\hat{s}\|}.
\end{equation}
Here, $s$ and $\hat{s}$ denote the original semantic embedding vector extracted from the source image and the recovered semantic embedding vector at the receiver, respectively \cite{fan2024semanticSim}.

{Image classification semantic loss:}
Similarly, some works adopt image classification semantic loss as the training objective for SRC, particularly in some wireless perception applications, where the correct recognition of visual element categories is more important than pixel-level reconstruction accuracy. The loss is formulated as
\begin{equation}
\mathcal{L}_{12} = -\sum y_{c} \log \hat{y}_{c},
\end{equation}
where $y_{c}$ and $\hat{y}_{c}$ are the ground-truth label and predicted posterior probability indicating whether the image element belongs to semantic category $c$, respectively \cite{wang2023semantic}.}

\minew{Other task-oriented losses: The encoder-decoder can be optimized end-to-end using losses from the specific task, such as bounding box regression and objectness score, or contrastive objectives for retrieval \cite{fan2024semanticSim}. These losses indirectly encourage the encoder to focus on task-relevant semantics.}

Ultimately, training SRC encoder-decoder architectures revolves around balancing semantic sufficiency with minimalism, ensuring that only necessary semantics are preserved, with robustness against channel-caused distortion and alignment with task-specific requirements.

\subsection{Implementation of Task-Oriented Modules}
Task-oriented modules serve as critical components that tailor SemCom systems to specific objectives and requirements. These modules are implemented differently across the three primary SemCom paradigms, each addressing distinct communication goals and constraints.

In SPC, task-oriented modules could be components to maintain fidelity and consistency throughout the transmission process. These modules typically include perceptual quality evaluators, semantic consistency validators, and adaptive error correction mechanisms. For instance, perceptual loss calculators that leverage pre-trained deep networks to measure semantic similarity between original and reconstructed content serve as essential task-oriented modules. In \cite{peng2024robust}, the authors demonstrate how LPIPS modules can be integrated into SPC systems to ensure perceptual consistency. Similarly, in \cite{xu2025semantic}, the semantic feature matching component is exploited to compare high-level representations extracted by convolutional NNs, thus helping preserve semantic integrity during transmission.
In SEC, task-oriented modules could be aesthetics scorers, style transfer components, and content enhancement modules that enrich the transmitted semantics with contextual or generative details. These modules leverage generative models to improve visual understanding and creativity of received content. For example, aesthetic quality assessment modules that predict visual appeal scores can guide the expansion process to generate more visually pleasing results. In \cite{liu2024cross}, the authors propose image quality measurement modules that can be integrated into SEC systems to evaluate and enhance aesthetic quality. 
\minew{In SRC, task-oriented modules could be area of interest predictors, saliency detectors, and task-relevance scorers that identify and prioritize the most important elements for specific applications. These modules enable efficient transmission by focusing on task-relevant information while discarding redundant details.} The authors in \cite{kang2022personalized} develop attention-based saliency modules that can predict regions of interest for various visual tasks, serving as fundamental components in SRC systems. Similarly, task-specific feature selectors that utilize learned importance weights to prioritize semantic features based on task requirements, as proposed in \cite{park2024transmit}, demonstrate effective implementation of refinement-focused modules.

In addition, ML models like reinforcement learning (RL) can be employed as task-oriented modules to dynamically adjust semantic transmission and resource allocation based on real-time conditions and performance feedback. These adaptive modules learn semantic transmission strategies through interaction with the communication environment, balancing multiple requirements such as bandwidth efficiency, latency, and reconstruction quality \cite{hua2024optimizing, hua2025bandwidth}. For example, in \cite{cheng2024wireless, cheng2025semantic}, the authors implement RL-based resource trade-off modules that adjust semantic computing and transmission in response to channel conditions and user requirements, demonstrating how intelligent task-oriented modules can optimize SemCom performance across varying scenarios. Similarly, multi-agent RL modules can coordinate multiple transceivers in distributed SemCom systems, as shown in \cite{shao2024semantic}, where collaborative learning enables efficient resource sharing and semantic understanding alignment.
\minew{Moreover, large language models can act as supervisory extension modules that manage multiple SemCom-Vision workflows, coordinating task intent, knowledge sharing, and module selection across different transceivers to support holistic and adaptive SPC, SEC, and SRC applications \cite{jiang2024large1, jiang2024large2, jiang2024large3,jiang2025visual,peng2024personalized,peng2025simac,jiang2025m4sc}.}

\subsection{Lessons Learned}
\minew{The design of the encoder–decoder reveals several critical insights that can guide future research and implementation of SemCom-Vision. Architecture choice plays a vital role, yet no single ML model can dominate across all SemCom-Vision categories with different communication goals. Hybrid designs that integrate complementary strengths of ML models often achieve superior performance. Equally important is the alignment of loss functions with communication goals. The SPC systems benefit from functions that preserve both semantic meaning and pixel fidelity, SEC systems require generative losses that encourage semantic richness while maintaining basic semantics, and SRC systems demand compact yet effective semantic representations. Moreover, task-oriented modules can further enhance adaptability by incorporating specialized components that can dynamically respond to varying task requirements.} 


\section{Knowledge for SemCom-Vision}
\label{sec:knowledge_management}

\begin{table*}[!t]
\centering
\caption{Overview of Knowledge Graph Structures and Potential Utilization in SemCom-Vision}
\label{Table_4_KG}
\begin{tblr}{
        colspec={Q[c,wd=1cm] Q[l,wd=2.8cm] Q[l,wd=2.5cm] Q[l,wd=5cm] Q[l,wd=4.6cm]},
        cell{5}{1} = {c = 5}{c,m},
        row{odd} ={bg = gray9},
}
\hline
    \textbf{Category}  & \textbf{Features} & \textbf{Typical CV Applications} & \textbf{Advantages and Limitations} & \textbf{Potential in SemCom-Vision}\\
\hline

    Triple store (TS)-based KGs 
    & RDF triples (subject, predicate, object), binary relations, ontology-based  
 
    & Image classification, object recognition, scene understanding, knowledge-driven visual reasoning  & Formal semantics, logic-based reasoning, standardized; \newline Rigid structure, text-centric, difficult to encode continuous visual features, triple explosion for rich visual contexts 
    & Ensures precise and task-relevant semantics in SRC, supports semantic association and inference in SEC, yet is less useful in SPC due to poor pixel-level and continuous feature representation. \\

    Property graph database (PGD)-based KGs   
    & Nodes/edges with arbitrary key-value properties, binary edges  
    & Visual relationship detection, scene graph construction, video understanding, visual knowledge retrieval  
    & Flexible schema, supports heterogeneous visual attributes, efficient traversal and querying;\newline Weaker formal logic support, reasoning often relies on external mechanisms  
    & Provides rich guidance for semantic extraction and reconstruction in SPC, enhances semantic association in SEC, and enables efficient task-oriented semantic filtering in SRC with low latency. \\

    Hyper-graph-based KGs
    & Nodes with hyperedges connecting multiple nodes (N-ary relations); hyperedge attributes  
    & Activity and context recognition, group interaction analysis, complex scene modeling  
    & Explicitly models high-order and multi-entity relations, captures complex contextual semantics;\newline Increased structural complexity, challenging real-time querying and storage  
    & Guides encoders to capture global and local semantics in SPC, supports context-aware semantic expansion in SEC, and enables composite semantic reasoning for efficient SRC. \\

\hline
    \textbf{Functional Extensions} & & & & \\
\hline

    Probabi-listic KGs  
    & Entities and relations associated with probabilities or distributions  
    & Uncertain visual perception, robust scene understanding, decision-making under noise
    & Models uncertainty, enables reasoning under noisy or unreliable conditions;\newline High computational complexity, challenging uncertainty estimation  
    & Prioritizes reliable semantics in SPC, supports plausible semantic inference in SEC, and enables risk-aware selection of task-relevant semantics in SRC to improve communication efficiency. \\
    Neural KGs  
    & Continuous embeddings, differentiable reasoning  
    & Visual-semantic embedding, cross-modal retrieval, neural-symbolic reasoning  
    & End-to-end optimization, adaptive reasoning, seamless integration with encoder–decoder models;\newline Reduced interpretability compared to symbolic KGs  
    & Enhances semantic preservation in SPC via learned priors, enables enriched semantic generation in SEC, and supports embedding-based task-oriented semantic selection in SRC under resource constraints. \\
    Multi-modal KGs  
    & Unified representation of text, image, audio, and video; nodes and edges carry multimodal features  
    & Multimodal scene understanding, vision–language tasks, cross-modal reasoning  
    & Rich cross-modal semantic context, complementary information fusion;\newline Potential scalability and latency issues if not carefully designed  
    & Maintains comprehensive semantic fidelity in SPC, enables contextually consistent semantic expansion in SEC, and supports efficient transmission of relevant cross-modal semantics in SRC. \\

\hline
\end{tblr}
\end{table*}

\minew{While encoder-decoder architectures serve as the backbone for semantic feature extraction and transmission, their performance heavily depends on the quality and relevance of the underlying knowledge representations \cite{liang2024generative}. Knowledge plays a central role in interpreting, organizing, and reasoning about semantics, and is capable of aiding semantic preservation, refinement, and expansion. This section reviews the research scope of KG in SemCom-Vision and potential improvements inspired by the works of KG in CV.}

\subsection{KG Structures in CV}
KG as a powerful tool for representing structured semantics in CV, organizing visual concepts into interconnected nodes and edges, thereby bridging the gap between low-level visual features and high-level semantic understanding. According to the model structure, KGs can be classified into three foundational types, i.e., triple store (TS), property graph databases (PGDs), and hypergraph, while some other special types are built as extensions or overlays on these, such as Probabilistic KGs, Multimodal KGs, and Neural KGs \cite{wang2017knowledge}.

\textbf{TS-Based KGs:}
The TS \cite{rohloff2007evaluation}, as a standardized and widely adopted structure, organizes knowledge into subject–predicate–object triples that conform to the resource description framework (RDF) \cite{lin2018knowledge}. Since each triple merely encodes a binary relation between two nodes, properties and complex semantics must be represented through additional triples.
\minew{
TS-based KGs are typically governed by ontologies, such as the web ontology language, enabling formal semantic reasoning. Additionally, these KGs support efficient rule-based inference via the semantic web rule language \cite{horrocks2004swrl}, and logical querying through SPARQL \cite{harris2013sparql}. While this structure allows rigorous rule expression and logical inference, it often leads to rigid representations and triple explosion when modeling high-dimensional data.
TS-based KGs have been explored in many CV research \cite{wang2017knowledge}. For example, the studies in  \cite{huang2024structure, liu2023revisiting, mi2024knowledge, pan2022contrastive} model semantic correspondences between visual and linguistic entities using TS-based KGs. By incorporating structured relational representations, these approaches enhance the semantic reasoning and cross-modal alignment capabilities of vision–language models such as CLIP and BLIP \cite{li2022blip}.}

\textbf{PGD-Based KGs:}
The PGD \cite{angles2020mapping} represents knowledge through nodes (entities), edges (relationships), and a set of properties that describe additional attributes of nodes/edges. \minew{This structure is commonly implemented in graph databases such as Neo4j and supports efficient traversal queries through languages like Cypher \cite{francis2018cypher}. PGD-based KGs offer richer modeling by attaching properties directly to both nodes and edges, unlike TS-based KGs that use extra nodes and edges for properties, allowing for more flexible and expressive data representation.} 
\minew{For instance, the authors in \cite{prabhu2015attribute} employ a PGD–based KG to represent image objects, their topological relationships, and associated properties like color, enabling efficient semantic understanding of images and accurate image similarity ranking. Similarly, in \cite{yang2023context}, nodes of the PGD are used to represent visual elements, edges capture inter-object relationships, while details of location and category are encoded as properties, thereby improving object detection and relational reasoning. In \cite{fang2020knowledge}, humans and devices are extracted from videos of construction sites as entities with corresponding relationships and properties to form a PGD-based KG, thus rapidly identifying and querying hazards.} 

\textbf{Hypergraph-Based KGs:}
\minew{Hypergraph-based KGs allow a single relation, i.e., hyperedge, to connect multiple entities simultaneously, thus explicitly modeling higher-order interactions. Instead of decomposing complex relations into multiple binary edges in TS-based KGs, a hyperedge represents a multi-way relationship with its own properties and semantics. This structure is particularly effective for capturing semantics in complex scenarios with many entities, at the cost of increased structural and querying complexity and fewer standardized implementations. In CV research, hypergraph-based KGs have proven effective for modeling complex semantics in visual data. 
For image segmentation, the authors of \cite{he2023multimodal} employ a hypergraph-based KG to parse the different visual modalities into irregular hypergraphs to mine semantic clues and generate robust mono-modal representations, thereby efficiently improving cross-modal compatibility when fusing diverse visual data semantics. In image recovery, a hypergraph empowered convolution module is exploited in \cite{fu2023continual} to better extract the nonlocal semantics with higher-order constraints on the data, thereby constructing a new backbone of the image to improve the image deraining performance.  In \cite{nguyen2025hyperglm}, a hypergraph-based KG is established to promote reasoning about multi-way interactions and higher-order relationships, effectively understanding complex semantics in diverse video scenes and helping video question answering, video generation, etc. } 

\textbf{Other Functional Extensions:}
\minew{Beyond the three fundamental KG structures, several specialized extensions, including \textit{probabilistic KGs}, \textit{neural KGs}, and \textit{multimodal KGs},  have emerged to enrich KG functionality. By solely or jointly incorporating these paradigms into foundational KG structures, KGs can be empowered with capabilities such as uncertainty-aware reasoning, continuous semantic representation learning, and cross-modal information integration.

\textit{Probabilistic KGs} extend deterministic knowledge representations by incorporating uncertainty quantification for entities and relationships through probability distributions \cite{kim2016probabilistic}. In challenging CV environments, where visual evidence is inherently uncertain, such KGs support principled reasoning under uncertainty through the integration of logical rules and probabilistic inference, thereby alleviating ambiguity induced by adverse sensing conditions \cite{choudhary2021probabilistic}. Moreover, by filtering low-confidence or irrelevant information, probabilistic KGs can improve query efficiency and accelerate downstream semantic processing \cite{lian2014keyword}. In CV research, probabilistic KGs have been employed for enhancing image understanding. In \cite{liu2024probabilistic}, uncertain object detections are refined through probabilistic entity representation model-based reasoning, where probability distributions over visual-semantic mappings enable robust generalization. 

\textit{Neural KGs} are essentially KGs that are empowered, enhanced, or learned using NN–based models, including graph NNs (GNNs) and knowledge representation learning (KRL) \cite{xie2016representation}. Unlike purely symbolic KGs, neural KGs enable differentiable reasoning and end-to-end optimization, which could be seamlessly integrated with ML-based image processing models \cite{werner2023knowledge}. Generally, in neural KGs, KRL can be used to learn vector embeddings of semantics, while GNNs like GraphSAGE can be exploited to further enhance both local and global semantic understanding via neighborhood aggregation \cite{hamilton2017inductive}. Moreover, the continuous nature of neural embeddings enables smooth interpolation between concepts and generalization to unseen visual entities, making them particularly valuable for zero-shot and few-shot learning scenarios \cite{chen2023zero},  which can significantly improve generalization and scalability in tasks like image recognition and question answering. The work of \cite{qiao2020neural} proposed a neural KG-based evaluator, where a powerful GNN is integrated with KG to achieve sophisticated feature extraction and combination. In \cite{nishihori2025transformer}, the authors introduce a neural KG by incorporating a transformer into the encoder of a contrastive learning framework, thereby enabling the model to dynamically incorporate related knowledge and facilitating more accurate correlations between entities and their properties.

\textit{Multimodal KGs} integrate information from diverse data modalities into unified knowledge representation \cite{song2024scene}. In multimodal KGs, each entity or relationship can be associated with multiple modal-specific semantics, such as latent space visual representations from VAEs and textual embedding from a Transformer \cite{lee2024multimodal}. With comprehensive information from different dimensions, Multimodal KGs are capable of supporting comprehensive scene understanding and alleviating ambiguity through cross-modal information integration. In \cite{wang2023tiva}, the authors propose TIVA-KG, a large-scale multimodal KG that explicitly aligns text, images, videos, and audio, enabling unified cross-modal representation and efficient reasoning to support visually grounded understanding tasks.
Similarly, the authors in \cite{lee2024multimodal,kannan2020multimodal} demonstrate how structured cross-modal entities and relations of multimodal KGs can enhance visual–language reasoning and complex inference beyond unimodal graph representations.}

\subsection{Existing Exploration of Knowledge in SemCom-Vision}
\minew{The role of knowledge in SemCom systems has evolved significantly, progressing from passive data repositories to active and reasoning-enabled components, which can be majorly classified as three stages.}

\textbf{Stage 1. Foundation Knowledge for Model Training: }
\minew{In its most fundamental form, knowledge serves as a large-scale annotated database that enables semantic encoder-decoders to learn meaningful representations during the training phase \cite{chaccour2024less}. These comprehensive databases, comprising labeled images,  contextual metadata, etc., allow ML models to internalize semantic representations. It guides the model learning but does not actively participate in the extraction, transmission, or reconstruction processes. For instance, in \cite{huang2022toward, xie2022task, dong2022semantic,tang2024contrastive, zhang2024scan,kang2022personalized}, the authors leverage large-scale image databases as training datasets to establish the foundational semantic understanding of their encoder-decoder architecture. While essential for model development, this stage treats knowledge purely as a pre-deployment resource, with no direct involvement in visual data transmission.}

\textbf{Stage 2. Structured Priors for Semantic Guidance:}
\minew{Advancing beyond passive training data, the second stage transforms raw knowledge into pre-processed, structured priors that actively inform semantic extraction, transmission, and reconstruction during operation. In this context, knowledge functions as a source of information prompts that provide additional contextual anchors for the semantics \cite{xing2024representation}. The majority of existing works exploit knowledge as the pre-processed priors to support semantic processing. In \cite{nam2024language}, pre-defined semantic categories and text guidance complement visual features, enabling more efficient semantic extraction and reconstruction. The work in \cite{park2025transmit} exploits structured datasets, including ImageNet with hierarchical taxonomies, COCO with object annotations and captions, and Visual Genome with scene graphs, as background knowledge to train semantic encoders while simultaneously providing categorical guidance for reconstruction. Similarly, the authors in \cite{lokumarambage2023wireless} employ the COCO dataset both for GAN training and as a common knowledge base that supplies object categories to aid image reconstruction in SEC, while the work of \cite{peng2025semantic} leverages structured knowledge to enhance semantic feature selection of SRC. 
Despite these advances, the knowledge usage in this stage remains fundamentally limited. The semantic understanding it provides is fixed, determined during training and embedded within model parameters. Moreover, it cannot adapt to novel scenarios or reason about missing information at inference time. The knowledge influences semantic processing but lacks explicit structure for active reasoning, dynamic query, or logical inference over relationships. }

\textbf{Stage 3. KGs for Reasoning-Driven Communication:}
\minew{In this stage, knowledge transitions from implicit statistical priors to explicit and structured KGs \cite{wang2017knowledge}. Beyond pre-processed datasets, KGs encode entities and their semantic relationships in queryable formats. This explicit structure enables SemCom systems to perform intelligent reasoning and inference, such as predicting missing visual elements, resolving ambiguities, and adapting transmission strategies based on relational context. 
Several pioneering works have begun integrating KGs into SemCom systems.  In \cite{liang2025knowledge}, a triple store (TS)-based KG framework is proposed to enhance the interpretability of text-oriented SemComs while mitigating semantic redundancy. The authors of \cite{zhou2023cognitive} exploit a TS-based KG to facilitate the SPC, where the semantic correction algorithm is proposed by mining the inference rules from the KG to avoid distortion of text semantics in the decoder. In \cite{xing2024representation}, the authors propose a TS-based KG in an SRC framework, where multimodal knowledge from the KG is integrated with visual semantics to enhance extraction and reconstruction accuracy.
In \cite{chaccour2024less, ren2024knowledge}, the authors introduce the potential of KG in visual data transmission from a research direction level. However, few existing studies have systematically investigated KG structures, construction, and utilization strategies tailored to different SemCom-Vision paradigms.}

\subsection{Potential of KGs in SemCom-Vision}
Inspired by CV studies and existing exploration in SemCom-Vision, the aforementioned KG structures exhibit different potential, motivating a systematic discussion of strengths and limitations in KG utilization for SemCom-Vision, as summarized in Table \ref{Table_4_KG}.
 
\textbf{TS-based KGs:} By transforming high-dimensional visual features into symbolic labels, TS-based KGs streamline semantic representations and the modeling of entities and relations, thereby supporting rigorous reasoning and inference in CV applications and also promising to deliver important information in SemCom-Vision.
However, TS KGs are inherently text-centric and are primarily designed for linguistic and symbolic knowledge, which makes them difficult to directly encode continuous visual features, such as spatial layout, texture, and detailed color, that are central to image understanding.
\minew{In SPC, where both pixel-level fidelity and semantic consistency are required, TS-based KGs exhibit inherent limitations. In SEC, TS-based KGs offer strong reasoning capabilities for semantically associating and inferring elements, but decomposing visual contexts into too many atomic triples for maintaining image richness may offset the efficiency gains targeted by SemCom-Vision systems. Nevertheless, rigorous reasoning and inference abilities of TS-based KGs can play a crucial role in SRC, where the primary emphasis is on identifying and conveying task-related semantics.
 }

\textbf{PGD-based KGs:}
\minew{PGD-based KGs hold significant potential for SemCom-Vision due to their structural flexibility and semantic richness. The multi-property node representation naturally accommodates heterogeneous visual features like entity color, while relationship edges can encode multiple semantic dimensions such as topology and functional relevance. Moreover, the PGD supports efficient subgraph querying and pattern matching, enabling efficient knowledge retrieval for semantic processing. }
\minew{In SPC, PGDs can provide encoders with rich guidance for semantic extraction, while the detailed properties associated with nodes and edges assist decoders in maintaining both pixel and semantic consistency during reconstruction. SEC can similarly benefit from the abundant information encoded in multi-property nodes and edges. Meanwhile, in SRC, the expressive query mechanisms of PGDs enable rapid semantic filtering and selection, facilitating the identification and prioritization of task-relevant semantics. Although PGD-based KGs do not support logic reasoning as rigorously as TS-based KGs, their high scalability, efficient computation, and low-latency query capabilities make them particularly attractive for SemCom-Vision applications with stringent latency requirements.}

\textbf{Hypergraph-Based KGs:}
\minew{Hypergraph-based KGs are particularly suitable for SemCom-Vision tasks that involve complex scenarios with multiple interacting entities. By allowing a single hyperedge to connect more than two nodes, they explicitly model higher-order relationships and multi-entity interactions.}
\minew{In SPC, hypergraph-based KGs can guide encoders to capture detailed semantics from global and local, ensuring both pixel-level fidelity and semantic consistency. For SEC, the multi-way relationships encoded in hyperedges enable decoders to infer or generate additional semantic details based on inter-object context, enhancing visual richness and creativity. In SRC, hypergraph-based KGs facilitate task-oriented reasoning by representing composite semantics, such as activities or scene-level semantics, allowing the selective transmission of only task-relevant information, thereby improving communication efficiency. Despite their expressive power, hypergraph KGs introduce increased structural complexity and pose challenges for efficient real-time querying.}

\textbf{Other Functional Extensions:} \minew{By introducing probabilistic, neural, and multimodal KGs as extensions, SemCom-Vision systems can model uncertainty, enable adaptive reasoning, and integrate cross-modal semantics, thereby extending the applicability and effectiveness of KGs across SPC, SEC, and SRC, as shown in Table \ref{Table_4_KG}.}

\minew{\textit{Probabilistic KGs} incorporate uncertainty into knowledge representations, which is crucial for visual data transmission over noisy or unreliable channels. By associating probability distributions with entities and relations, they enable reasoning under uncertainty.}
\minew{For SPC, probabilistic KGs can prioritize reliable semantics to preserve fidelity. In SEC, they support the inference of plausible additional semantics under uncertain conditions. Moreover, probabilistic reasoning can facilitate the risk-aware selection of task-relevant semantics in SRC, thereby maximizing efficiency while minimizing the transmission of low-confidence or redundant information. Despite the benefits, the computational complexity and accurate uncertainty estimation of probabilistic KGs remain key challenges in real-time tasks.}

\minew{\textit{Neural KGs} integrate symbolic knowledge with continuous, differentiable embeddings, enabling end-to-end optimization with encoder-decoders. This makes them highly adaptive to dynamic SemCom-Vision scenarios.}
\minew{In SPC, neural KGs can provide priors that enhance semantic preservation by aligning detailed visual features with latent space semantics. For SEC, differentiable reasoning of neural KGs can support the generation of enriched semantics. In SRC, embedding-based representations allow task-oriented semantic selection, improving efficiency in scenarios with limited bandwidth or computational resources. Their main limitation lies in reduced interpretability compared with symbolic KGs, which may be critical in certain applications.}

\minew{\textit{Multimodal KGs} unify cross-modal knowledge into a single structured representation, providing comprehensive semantic context.}
\minew{In SPC, multimodal KGs can guide the extraction and reconstruction of comprehensive semantics across modalities to maintain high fidelity. Moreover, by leveraging complementary modal information, they enable the generation of enriched and contextually consistent semantic details in SEC, while transmitting only the most relevant and accurate cross-modal information in SRC. However, when applied to latency-sensitive SemCom-Vision scenarios, multimodal KGs must be carefully designed to limit graph scale and structural complexity, ensuring low-overhead and efficient reasoning.}

\section{Potential Applications} \label{sec:applications}

SemCom provides a pathway toward more efficient information transmission. By aligning transmitted content with different tasks and goals, the SPC, SEC, and SRC offer transformative potential across a wide range of intelligent applications. In this section, we explore key case studies where SemCom plays a critical role.

\subsection{Digital Twins}
Digital twins represent sophisticated digital replicas of physical systems, processes, or entities that enable continuous monitoring, analysis, and decision-making through bidirectional data synchronization \cite{alkhateeb2023real}. To create accurate virtual replicas of physical counterparts and dynamically maintain them, numerous data with different modalities could necessitate being collected and transmitted from various sensors \cite{mihai2022digital}. In this context, transmitting all data indiscriminately, especially when much of it remains static and contributes minimally to synchronization, can further exacerbate bandwidth constraints. Given that digital twins also inherently serve as databases, managing the continuously accumulating and massive volumes of data becomes increasingly challenging \cite{khan2022digital}.

SemCom can be exploited as a potential solution, especially SPC and SRC. SPC enables efficient transmission by conveying only the semantics of raw data between physical systems and digital replicas, thereby alleviating communication resource demands. Meanwhile, SRC can analyze and prioritize the semantics relevant to necessary synchronization, such as anomalies or system failures \cite{liu2025anomaly}. Moreover, the use of KGs within SemCom can support semantic-level database management for digital twins, orchestrate data for different tasks and contextual interpretations, thus enabling more effective data selection, analysis, and cyber-physical decision-making. Some existing works \cite{ thomas2023causal, tang2024uav, austin2020architecting} exploit semantic models in digital twins. In \cite{thomas2023causal}, the authors propose an SPC-based digital twin synchronization framework,  where semantics are extracted from the raw sensed data transmission, followed by the recovery of semantic information on the edge server. Reference \cite{tang2024uav} proposes an unmanned aerial vehicle-assisted edge computing system empowered with SRC, which reduces synchronization latency of digital twins by transmitting the desired semantics. Reference \cite{austin2020architecting} introduces a semantic empowered digital twin for a smart city, where semantic-based KG and ontology are exploited in the digital twin to provide complementary and supportive roles in the collection and processing of data, identification of events, and automated decision-making. 

\subsection{Metaverse}
The metaverse envisions an immersive digital environment where users interact through avatars, real-time rendering, and multimodal communication. It integrates technologies such as augmented reality, virtual reality, extended reality, digital avatars, spatial computing, and natural language processing to enable seamless interactions across physical and virtual spaces \cite{tang2022roadmap}. Similar to digital twins, maintaining responsiveness in the metaverse requires transmitting vast amounts of heterogeneous data, such as audio, video, gestures, gaze, and spatial coordinates, which can result in significant communication overhead \cite{khan2024metaverse}. Additionally, to ensure immersion, the metaverse has an ultra-low latency requirement, which further brings challenges \cite{yu2023asynchronous}. Furthermore, the metaverse requires transmitting sensitive information like facial features and eyeball details, which raises privacy concerns \cite{wang2022survey}. 

SemCom presents a promising solution to the data deluge in the Metaverse by prioritizing meaning over raw data fidelity. In particular, SRC can be used to prioritize and transmit only task-specific semantics, such as user intentions, emotional cues, focal objects, or key object interactions,  while simplifying less important elements, for example, a meaningless wall background, thereby dramatically reducing bandwidth consumption. Moreover, it enhances privacy protection by recognizing and omitting the transmission of sensitive elements, such as facial features. Even if the corresponding semantics are eavesdropped during transmission, they are difficult to reconstruct as the original information without the matching decoder.
Furthermore, the SEC enables personalized services by fusing different data and increasing the depth, contextual diversity, and generative richness according to diversified user needs.
Recent research \cite{wang2023semantic2,lin2023unified,chen2023trustworthy} efforts reflect these directions. In \cite{wang2023semantic2}, an SRC-aware data transmission framework is proposed for the metaverse, which prioritizes important sensor data, thus reducing the amount of data significantly and lowering the storage and transmission costs. In \cite{lin2023unified}, the authors utilize SEC to dynamically process multimodal input data (speech, eye movement, hand tracking) for real-time interaction in the metaverse, and ensure goal-oriented high-quality content generation to improve immersion from both communication and content perspectives. \minew{Moreover, the work of \cite{chen2023trustworthy} explores the integration of SemCom into Metaverse platforms to transmit private multi-modal data and exhibits its potential in guarding against security breaches and privacy leakage.} Such integrations show the potential of combining SemCom with semantic knowledge to create scalable, intelligent, and user-centric Metaverse systems.

\subsection{Wireless Perception}
Wireless perception systems rely on the acquisition, processing, and transmission of multimodal data to enable real-time perception, recognition, and decision-making in diverse environments \cite{xu2023edge}. These systems generate and exchange heterogeneous data sources, such as radio frequency (RF) signals, radar echoes, acoustic waves, and visual measurements \cite{du2023rethinking}. Transmitting such high-dimensional data over wireless channels, often under resource constraints, dynamic channel conditions, and diverse latency requirements, presents significant challenges in terms of bandwidth, energy efficiency, and reliability \cite{sagduyu2024joint}.

SemCom is promising to address these issues by enabling content-aware and task-oriented data transmission. SRC can dynamically identify and transmit only the features crucial for urgent sensing tasks, such as sudden target motion or abnormal activity detection. By selectively prioritizing high-impact semantics, the system reduces transmission overhead while ensuring timely responses to critical events.
Meanwhile, SEC can integrate semantics from multiple visual modalities to fuse a comprehensive semantic representation, thereby ensuring accurate sensing while ensuring more efficient use of limited spectrum resources. Moreover, SEC can be leveraged to reconstruct or infer missing semantics when sensing signals are incomplete or distorted. In scenarios such as sensor occlusion or partial sensor failure, SEC can exploit generative models and contextual priors to recover essential sensing information.   
Recent works of \cite{lv2024importance,gimenez2024semantic,wang2025deep} have started to explore such SemCom-based solutions in wireless perception. For instance, the work of \cite{lv2024importance} proposes importance-aware SRC for selective feature transmission, reducing bandwidth while preserving sensing accuracy. Similarly, in \cite{gimenez2024semantic}, the authors exploit SRC encoders to process data collected by distributed roadside units and identify important semantics based on segmentation, which greatly improves the performance in terms of maximum supported load and latency. In \cite{wang2025deep}, the authors exploit SEC to reconstruct degraded sensing signals, demonstrating improved robustness and semantic completeness under channel impairments.
Through the adoption of SemCom, wireless perception systems can achieve more intelligent, adaptive, and resource-efficient operation, paving the way for robust sensing in dynamic and resource-constrained environments.

\section{Conclusion} \label{sec:conclusion}
In this survey, we have systematically reviewed the emerging field of SemCom-Vision from an ML perspective. We have provided a clear explanation of the foundation and basics of the SemCom and elaborated on the classification of SemCom-Vision based on semantic quantization schemes and communication goals. Our analysis of major encoder-decoder architectures alongside ML models and training strategies offers guidance for designing SemCom-Vision across diverse tasks and goals. Additionally, we have introduced the knowledge structure and utilization for further enhancing semantic reasoning and processing. Moreover, the discussion of practical applications in fields of digital twins, metaverse, and wireless perception showcases the real-world potential of SemCom-Vision. 
By providing this structured overview of SemCom-Vision and systematically identifying critical research avenues, this survey serves as a foundational reference for both researchers and practitioners. It aims to inspire and guide future work to effectively integrate CV and communication engineering, fostering the development of more intelligent, efficient, and robust SemCom-Vision systems.

  \section*{Acknowledgments}

  This work is supported by the Engineering and Physical Sciences Research Council, under grant EP/Z533221/1 (TransiT: Digital Twinning Research Hub for Decarbonising Transport).
 


\bibliographystyle{IEEEtran}
\bibliography{IEEEabrv, SemSurvey.bib}

\begin{IEEEbiography}[{\includegraphics[width=1in,height=1.25in,clip,keepaspectratio]{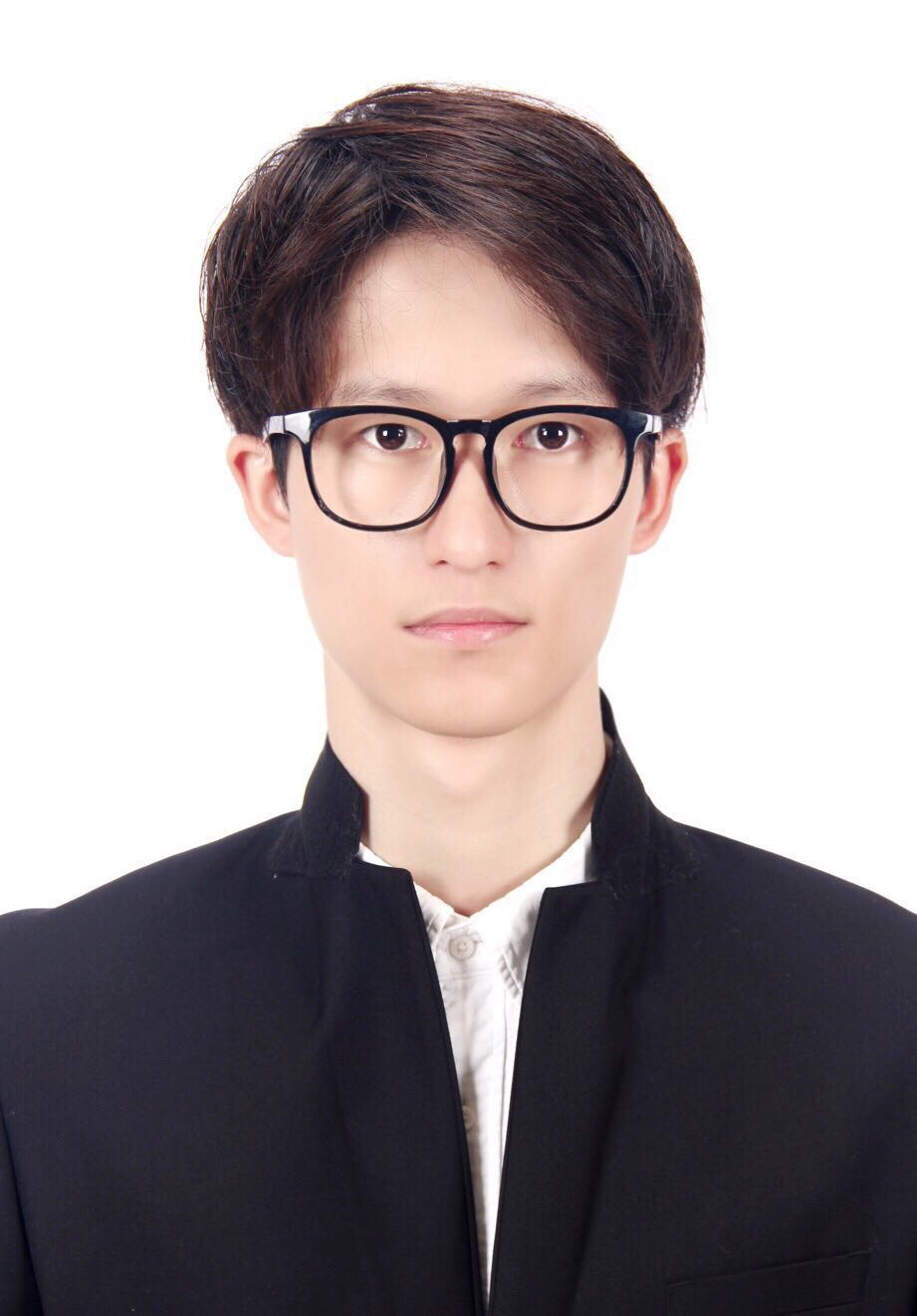}}]{Runze Cheng}
    (Member, IEEE) received the Ph.D. degree in Electrical and Electronic Engineering from the James Watt School of Engineering, University of Glasgow, U.K. in 2023. He is currently a Postdoctoral Research Associate with the Communications, Sensing, and Imaging Research Group, University of Glasgow, Glasgow, U.K. His research interests include intelligent resource management, semantic communication, space-air-ground integrated networks, and digital twins.  
\end{IEEEbiography}

\begin{IEEEbiography}[{\includegraphics[width=1in,height=1.25in,clip,keepaspectratio]{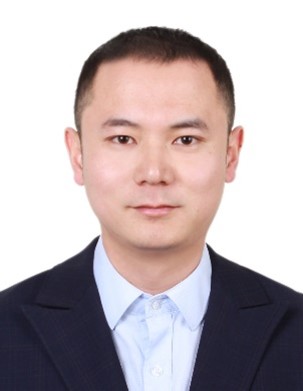}}]{Yao Sun}
    (Senior Member, IEEE) is currently a senior lecturer with the James Watt School of Engineering, University of Glasgow, Glasgow, UK. Dr. Sun has extensive research experience in wireless communication area. He has won the IEEE IoT Journal Best Paper Award 2022, and IEEE Communication Society of TAOS Best Paper Award in 2019 ICC. His research interests include intelligent wireless networking, SemCom and wireless blockchain system. 
\end{IEEEbiography}

\begin{IEEEbiography}[{\includegraphics[width=1in,height=1.25in,clip,keepaspectratio]{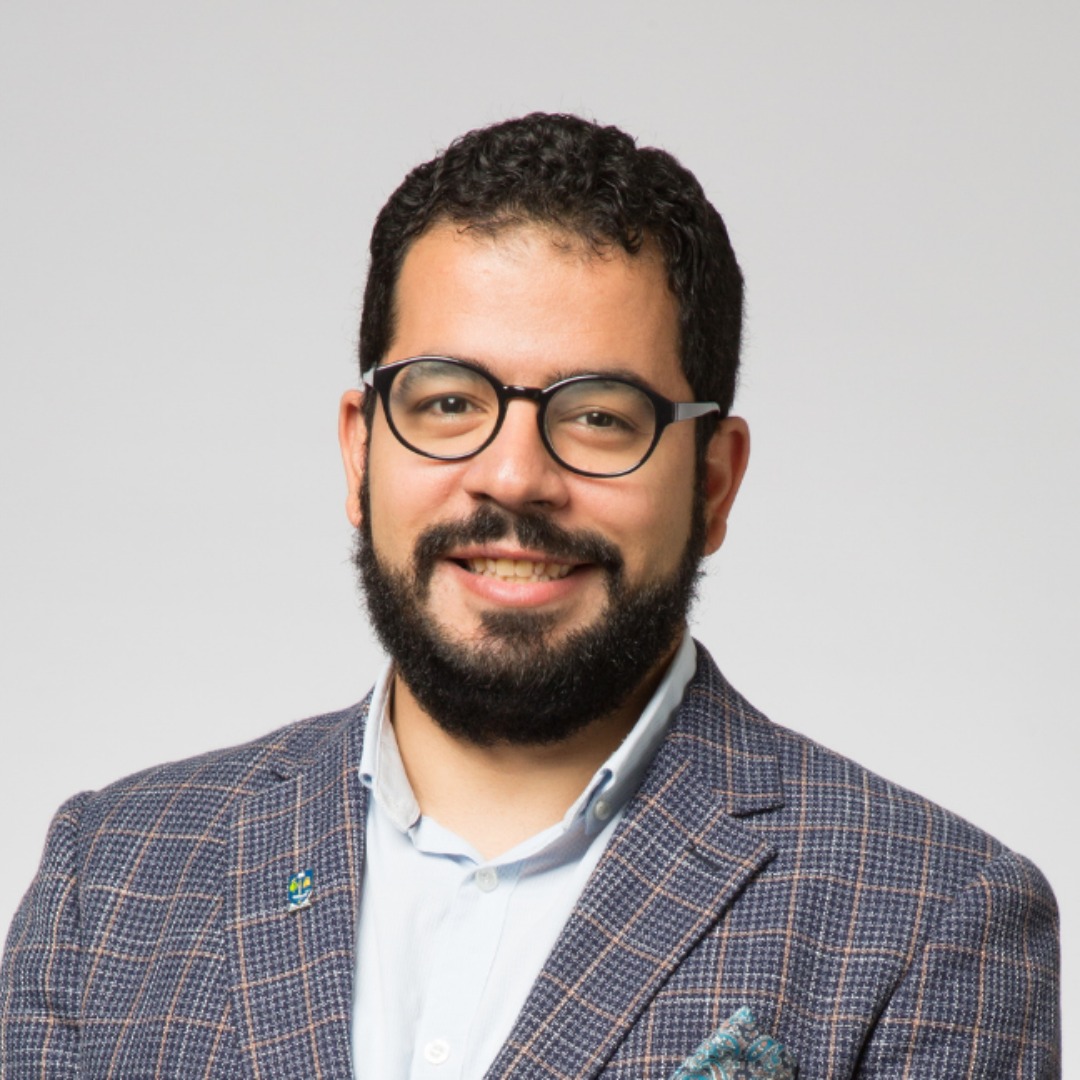}}]{Ahmad Taha}
    (Senior Member, IEEE) is a Lecturer in Autonomous Systems and Connectivity at the James Watt School of Engineering, University of Glasgow, with over a decade of experience across leading global higher education institutions. His research explores the energy nexus through cyber‑physical systems, leveraging digital twins and IoT technologies to enable decarbonisation. He is a member of the Glasgow Retrofit Advisory Group, advising on digitalisation to decarbonise housing.
    Dr Taha leads the connectivity research agenda associated with the multi‑million‑pound EPSRC-funded TransiT hub, supporting large‑scale infrastructure integration through advanced sensing, communications, and cyber‑physical architectures.
    He has authored and co‑authored publications in reputable venues and has contributed as a Principal Investigator and Co‑Investigator to research grants exceeding £20 million. He has been recognised by the Royal Academy of Engineering through the Global Talent scheme. Dr Taha is an inaugural member of the UK Young Academy, an academic advisor to the Commonwealth Scholarship Commission, a Fellow of Advance HE (FHEA), and a member of the International Science Council Global Roster of Experts.
\end{IEEEbiography}

\begin{IEEEbiography} [{\includegraphics[width=1in,height=1.25in,clip,keepaspectratio]{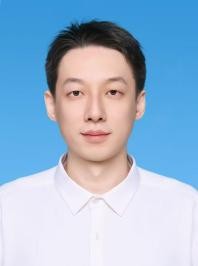}}]{Xuesong Liu}(Student Member, IEEE) is pursuing his Ph.D. degree at the University of Glasgow. He has won the Philip Merlin Best Paper Award, his master's thesis was awarded the distinction of best graduation thesis in Newcastle University 2022. His research interests include privacy preserving, semantic communication, Secure Wireless Communication.
\end{IEEEbiography}

\begin{IEEEbiography} [{\includegraphics[width=1in,height=1.25in,clip,keepaspectratio]{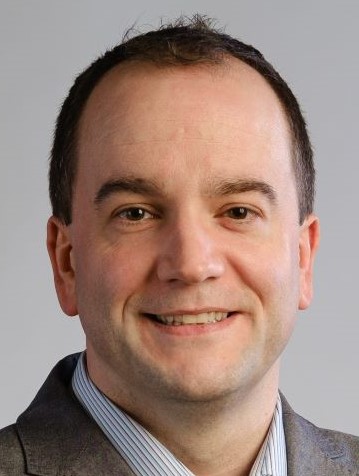}}]{David Flynn}  
    (Senior Member, IEEE) received the B.Eng. degree (Hons.) in electrical and electronic engineering, the M.Sc. degree (Hons.) in microsystems, and the Ph.D. degree in microscale magnetic components from Heriot-Watt University in 2002, 2003, and 2007, respectively. He is currently a Professor of cyber physical systems with the University of Glasgow. He is also the Co-Founder of U.K.’s EPSRC National Centre for Energy Systems Integration and U.K. Robotics and Artificial Intelligence Hub for Offshore Energy Asset Integrity Management.
\end{IEEEbiography}

\begin{IEEEbiography} [{\includegraphics[width=1in,height=1.25in,clip,keepaspectratio]{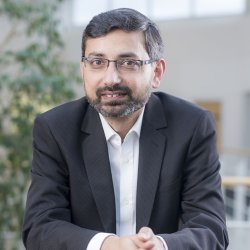}}]{Muhammad Ali Imran}  
    (Fellow, IEEE) eceived the M.Sc. (Hons.) and Ph.D. degrees from Imperial College London, London, U.K., in 2002 and 2007, respectively. He is currently the Dean of the University of Glasgow and UESTC; the Head of the Autonomous Systems and Connectivity Division; and a Professor of communication systems with the James Watt School of Engineering, University of Glasgow, Glasgow, U.K. He is an Affiliate Professor with The University of Oklahoma, Norman, OK, USA, and 5G Innovation Centre, University of Surrey, Guildford, U.K. He is leading research with the University of Glasgow for Scotland 5G Centre. He has more than 20 years of combined academic and industry experience with several leading roles in multi-million pound funded projects, working primarily in the research areas of cellular communication systems. He was a recipient of the Award of Excellence in recognition of his academic achievements, conferred by the President of Pakistan. He was also a recipient of the IEEE Comsoc’s Fred Ellersick Award 2014, the FEPS Learning and Teaching Award 2014, the Sentinel of Science Award 2016, and ten patents.
\end{IEEEbiography}

\end{document}